\documentclass[11pt, onecolumn]{IEEEtran} 

\usepackage{pdfsync}
\usepackage{amsmath}
\usepackage{amssymb}
\usepackage{amsfonts}
\usepackage{epsfig}
\usepackage{subfigure}
\usepackage{graphicx}
\usepackage{cite}
\usepackage{calc}
\usepackage{color}
\usepackage[all]{xy}
\usepackage{bm}
\title{Linear Coding Schemes for the Distributed Computation of Subspaces}

\newtheorem{thm}{Theorem}

\newtheorem{lem}[thm]{Lemma}
\newtheorem{cor}[thm]{Corollary}
\newtheorem{note}{Remark}
\newtheorem{ex}{Example}

\newcommand{\beq}{\begin{equation}}
\newcommand{\eeq}{\end{equation}}

\newcommand{\bea}{\begin{eqnarray}}
\newcommand{\eea}{\end{eqnarray}}
\newcommand{\bean}{\begin{eqnarray*}}
\newcommand{\eean}{\end{eqnarray*}}
\newcommand{\bit}{\begin{itemize}}
\newcommand{\eit}{\end{itemize}}
\newcommand{\ben}{\begin{enumerate}}
\newcommand{\een}{\end{enumerate}}
\newcommand{\blem}{\begin{lem}}
\newcommand{\elem}{\end{lem}}
\newcommand{\bthm}{\begin{thm}}
\newcommand{\ethm}{\end{thm}}
\newcommand{\bpf}{\begin{proof}}
\newcommand{\epf}{\end{proof}}

\newcommand{\bprf}{{\em Proof: }}

\newcommand{\eproof}{\hfill $\Box$} \author{ V. Lalitha, N. Prakash, K. Vinodh, P. Vijay Kumar and S. Sandeep Pradhan \thanks{V. Lalitha, N. Prakash, K. Vinodh and P. Vijay Kumar are with the Department of ECE, Indian Institute of Science, Bangalore,
560 012 India (email: \{lalitha, prakashn, kvinodh, vijay\}@ece.iisc.ernet.in).}
\thanks{S. Sandeep Pradhan is with the Department of EECS, University of Michigan, Ann Arbor, MI 48109, USA
(email: pradhanv@eecs.umich.edu).}
\thanks{The results in this paper were presented in part at the $49^{\text{th}}$ Allerton
Conference on Communications, Control and Computing, Monticello, IL,
Sept. 2011 \cite{LalPraVinKumPra}. }
\thanks{The work of P. Vijay Kumar was supported in part by the US National Science Foundation under Grant 0964507.  The work of V. Lalitha is supported by a TCS Research Scholarship and and the work of K. Vinodh is supported by Microsoft Corporation and Microsoft Research India under the Microsoft Research India PhD Fellowship Award.} }

\begin{document}

\maketitle

\thispagestyle{empty}
\begin{abstract}
Let $X_1, ..., X_m$  be a set of $m$ statistically dependent sources over the common alphabet $\mathbb{F}_q$, that are linearly independent when considered as functions over the sample space.  We consider a distributed function computation setting in which the receiver is interested in the lossless computation of the elements of an $s$-dimensional subspace $W$ spanned by the elements of the row vector $[X_1, \ldots, X_m]\Gamma$ in which the $(m \times s)$ matrix $\Gamma$  has rank $s$.  A sequence of three increasingly refined approaches is presented, all based on linear encoders.

The first approach uses a common matrix to encode all the sources and a Korner-Marton like receiver to directly compute
$W$. The second improves upon the first by showing that it is often more efficient to compute a carefully chosen superspace
$U$ of $W$.  The superspace is identified by showing that the joint distribution of the $\{X_i\}$ induces a unique
decomposition of the set of all linear combinations of the $\{X_i\}$, into a chain of subspaces identified by a normalized
measure of entropy. This subspace chain also suggests a third approach, one that employs nested codes.   For any joint
distribution of the $\{X_i\}$ and any $W$, the sum-rate of the nested code approach is no larger than that under the
Slepian-Wolf (SW) approach. Under the SW approach, $W$ is computed by first recovering each of the $\{X_i\}$.  For a large
class of joint distributions and subspaces $W$, the nested code approach is shown to improve upon SW.   Additionally, a class
of source distributions and subspaces are identified, for which the nested-code approach is sum-rate optimal.

\end{abstract}


\section{Introduction} \label{sec:intro}
In \cite{KorMar}, Korner and Marton consider a distributed source coding problem with two discrete memoryless binary
sources $X_1$ and $X_2$ and a receiver interested in recovering their modulo-two sum $Z = X_1 + X_2 \bmod 2$. An obvious
approach to this problem would be to first recover both $X_1$ and $X_2$ using a Slepian-Wolf encoder \cite{SleWol} and then
compute their modulo-two sum thus yielding a sum-rate of $H(X_1,X_2)$. Korner and Marton present an interesting, alternative
approach in which they first select a $(k \times n)$ binary matrix $A$ that is capable of efficiently compressing $\bold{Z} =
 \bold{X}_1 \ + \ \bold{X}_2 \bmod 2$, where ${\bf X}_1, {\bf X}_2$ and ${\bf Z}$ correspond to i.i.d. $n$-tuple realizations of $X_1, X_2$ and $Z$ respectively. The two sources then transmit $A\bold{X}_1$ and $A\bold{X}_2$ respectively.   The receiver
first computes $A\bold{X}_1 \ + \ A\bold{X}_2 \ = \ A\bold{Z} \bmod 2$ and then recovers $\bold{Z}$ from $A\bold{Z}$. Since
optimal linear compression of a finite field discrete memoryless source is possible \cite{HanKob}, the compression rate
$\frac{k}{n}$ associated with $A$ can be chosen to be as close as desired to $H(Z)$, thereby implying the achievability of
the sum rate $2H(Z)$ for this problem. For a class of symmetric distributions, it is shown that this rate not only improves
upon the sum rate, $H(X_1, X_2)$, incurred under the Slepian-Wolf approach, but is also optimum.

In this paper, we consider a natural generalization of the above problem when there are more than two statistically
dependent sources and a receiver interested in recovering multiple linear combinations of the sources. Our interest is in
finding achievable sum rates for the problem and we restrict ourselves to linear encoding in all our schemes.

\subsection{System Model}  \label{sec:system_model}

Consider a distributed source coding problem involving $m$ sources $X_1, ..., X_m$ and a receiver that is interested in the
lossless computation (i.e., computation with arbitrarily small probability of error) of a function of these sources. All
sources are assumed to take values from a common alphabet, the finite field $\mathbb{F}_q$ of size $q$.  The
sources are assumed to be memoryless and possessing a time-invariant joint distribution given by $P_{X_1\ldots X_m}$.   We
will assume this joint distribution $P_{X_1\ldots X_m}$ to be ``linearly non-degenerate'', by which we mean that when
the random variables $\{X_i, 1 \leq i \leq m\}$ are regarded as functions over the sample space $\Omega$, they are
linearly independent, i.e.,
\bean
\sum_{i=1}^m a_i X_i(\omega) & = & 0 , \ \ \text{ all } \omega \in \Omega, \ \ a_i \in \mathbb{F}_q,
\eean
iff $a_i=0$, all $i$.    For simplicity in notation, we will henceforth drop $\omega$ in the notation.  By identifying the
linear combination $\sum_{i=1}^ma_iX_i$ with the vector $[a_1\ a_2 \ \cdots a_m]^T$, we see that the vector space $V$ of all
possible linear combinations of the $\{X_i\}$ can be identified with $\mathbb{F}_q^m$.

The function of interest at the receiver is assumed to be the set of $s$ linear combinations $\{Z_i \mid i=1,2,\ldots, s\}$
of the $m$ sources given by:
\begin{eqnarray} \label{eq:prob_statement}
[Z_1, \ldots, Z_s] & = & [X_1, \ldots, X_m]\Gamma,
\end{eqnarray}
in which $\Gamma$ is an $(m \times s)$ matrix over $\mathbb{F}_q$ of full rank $s$ and where matrix multiplication is over
$\mathbb{F}_q$. Note that a receiver which can losslessly compute $\{Z_i, \ i = 1, \ldots, s\}$ can also compute any linear
combination $\sum_{i=1}^{s} \beta_iZ_i, \ \beta_i \in \mathbb{F}_q$, of these random variables.  The set of all linear
combinations of the $\{Z_i, \ i = 1, \ldots, s\}$ forms a subspace $W$ of the vector space $V$, which can be identified with
the column space of the matrix $\Gamma$. This explains the phrase `computation of subspaces ' appearing in the
title.  Throughout this paper, we will interchangeably refer to the $\{X_i, i = 1 \ldots, m\}$ as random variables (when they
refer to sources) and as vectors (when they are considered as functions on the sample space). We will also write $V = <X_1, \ldots, X_m>$ to
mean that $V$ is generated by the vectors $\{X_i, \ 1 \leq i \leq m\}$.  Similarly with other random variables and their
vector interpretations.

\begin{figure}[h!]
\begin{center}
\epsfig{figure=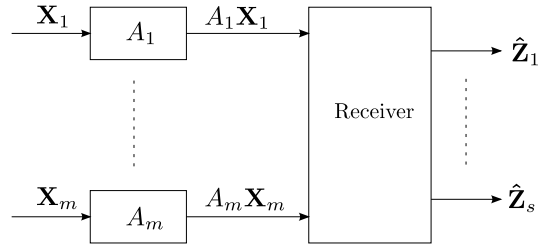,width=7cm}
\caption{The common structure for all three approaches to subspace computation.}
\label{fig:prob_statement} \end{center}
\end{figure}

{\textit{Encoder}}: All the encoders will be linear and will operate on $n$-length i.i.d. realizations, of the corresponding
sources $X_i,  1 \leq i \leq m$. Thus the $i$th encoder will map the $n$-length vector $\mathbf{X}_i$ to $A_i^{(n)} {\bf X}_i$ for some $(k_i \times n)$ matrix $A_i^{(n)}$ over $\mathbb{F}_q$.  The rate of the $i^{\text{th}}$ encoder, in bits per symbol, is thus given by
\bea
R_i^{(n)} & = & \frac{k_i}{n} \log q \ .
\eea

\vspace*{.1in}

{\textit{Receiver}}: The receiver is presented with the problem of losslessly
recovering the $\{{\bf Z}_i, 1 \leq i \leq s\}$, which are $n$-length extensions of the random variables $\{Z_i, 1 \leq i
\leq s\}$ defined in \eqref{eq:prob_statement}, from the $\{A_i{\bf X}_i, 1 \leq i \leq m\}$.   Let ${\bf W}$ be the space
spanned by the $\{{\bf Z}_i, 1 \leq i \leq s\}$.  Then lossless recovery of the $\{{\bf Z}_i\}$ amounts to lossless
computation of the subspace ${\bf W}$.  Thus in the present notation, ${\bf W}$ is to $W$ as ${\bf Z}_i$ is to $Z_i$.

For $1 \leq i \leq s$, let  $\widehat{\bold{Z}}_i$ denote the receiver's estimates of $\bold{Z}_i$.  We will use $P_e^{(n)}$
to denote the probability of error in decoding, i.e.,
\begin{eqnarray}
P_e^{(n)} & = & P\left( (\bold{Z}_1 \ \ldots \ \bold{Z}_{s}) \neq (\widehat{\bold{Z}}_1 \ \ldots \ \widehat{\bold{Z}}_{s})
\right).
\end{eqnarray}

\vspace*{.1in}

{\textit{Achievability}}: A rate tuple $(R_1, \ldots, R_m)$ is said to be \textit{achievable}, if for any $\delta > 0$, there
exists a sequence of matrix encoders $\left\{\left(A_1^{(n)}, \ldots, A_m^{(n)}\right)\right\}_{n = 1}^{\infty}$ (and
receivers) such that $R_i^{(n)} \leq R_i  + \delta, 1 \leq i \leq m$, for sufficiently large $n$, and $\lim_{n
\rightarrow \infty}P_e^{(n)} = 0$. A \textit{sum rate} $R$ will be declared as being achievable, whenever $R=\sum_{i =
1}^{m}R_i$ for some achievable rate tuple $(R_1, \ldots, R_m)$. By \textit{rate region} we will mean the closure of set of
achievable rate $m$-tuples. In situations where all encoders employ a common matrix encoder, we will then use the term
\textit{minimum symmetric rate} to simply mean the minimum of all values $R$ such that the symmetric point $(R_1=R, \ldots,
R_m=R)$ lies in the rate region.

\subsection{Our Work} \label{sec:our_work}

In this paper, we present three successive approaches to the subspace computation problem along with explicit characterization of the corresponding achievable sum rates. As illustrated (Fig.
\ref{fig:prob_statement}), all three approaches will use linear encoders, but will differ in their selection of encoding matrices $A_i$.  We provide an overview here, of the three approaches along with a brief description of their achievable sum rates, and some explanation for the relative performance of the three approaches.  Details and proofs appear in subsequent sections.

\vspace{0.1in}

\begin{figure}[h!]
\begin{center}
\includegraphics[width=8cm]{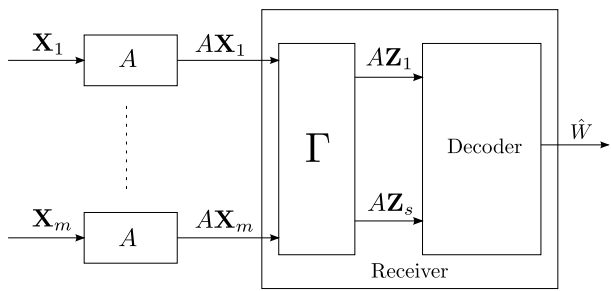}
\caption{The Common Code approach.}
\label{fig:CC_approach} \end{center}
\end{figure}

\textit{Common Code (CC) approach}: Under this approach, the encoding matrices of all the sources are assumed to be
identical, i.e., $A_i = A, \ 1 \leq i \leq m$.  It is also assumed, that the receiver decodes $[\bold{Z}_1 \ \ldots \
\bold{Z}_{s}]$ by first computing (Fig. \ref{fig:CC_approach})
\bea
\left[ A\bold{Z}_1 \ \  \cdots \ \ A\bold{Z}_s \right]   & = &  \left[ A\bold{X}_1 \ \ \cdots
\ \ A\bold{X}_m \right] \Gamma,
\eea
and thereafter processing the $\{A\bold{Z}_i\}$.
Thus the CC approach could be regarded as the analogue of the Korner-Marton approach for the modulo-two sum of two sources described earlier.  The minimum symmetric rate under this approach is characterized in Theorem \ref{thm:CC}.

\vspace{0.1in}

\textit{Selected Subspace Approach}: It turns out interestingly, that compression rates can often be improved by
using the CC approach to compute a larger subspace $U \subseteq V$ that contains the desired subspace $W$, i.e., $W \subseteq
U$. We will refer to this variation of the common code approach under which we compute the superspace $U$ of $W$, as the
\textit{Selected Subspace (SS) approach}. Thus when we speak of the SS approach, we will mean the selected-subspace variation
of the common-code approach. We will present in the sequel (Theorem \ref{thm:chain_SSA}), an analytical means of determining
for a given subspace $W$, the best subspace $U \supseteq W$ upon which to apply the CC approach.   This is accomplished by
showing (Theorem \ref{thm:chain_NCE}) that the joint distribution of the $\{X_i\}$ induces a unique decomposition of the
$m$-dimensional space $V$ into a chain of subspaces identified by a normalized measure of entropy.   Given this
subspace chain, it is a simple matter to determine the optimum subspace $U$ containing the desired subspace $W$.
\vspace{0.1in}

\begin{ex} \label{ex:opt_dist_intro}
Consider a setting where there are $4$ sources, $X_1, \ldots, X_4$ having a common alphabet $\mathbb{F}_2$, whose joint
distribution is described as follows:
\begin{equation}
\left[\begin{array}{c} X_1 \\ X_2 \\ X_3 \\ X_4 \end{array}\right] \ = \
\left[ \begin{array}{cccc} 1 & 1 & 1 & 1 \\
0 & 1 & 1 & 1 \\
0 & 0 & 1 & 1 \\
0 & 0 & 0 & 1
\end{array}\right] \left[\begin{array}{c} Y_1 \\ Y_2 \\ Y_3 \\ Y_4 \end{array}\right], \label{eq:ex_intro_temp1}
\end{equation}
where $\{Y_i\}_{i=1}^{4}$ are independent random variables such that $Y_1, Y_2 \sim \text{Bernoulli}(p_1), Y_3 \sim
\text{Bernoulli}(p_2), Y_4 \sim \text{Bernoulli}(\frac{1}{2}), 0 < p_1 < p_2 < \frac{1}{2}$.  Assume that
the receiver is interested in decoding the single linear combination $Z = X_1 + X_2 + X_3+ X_4 \mod 2$.  In the subspace
notation, this is equivalent to decoding the one dimensional space $W = <Z>$. The CC approach would choose the common
encoding matrix $A$ so as to
compress $Z$ to its entropy, $H(Z)$, thus yielding a sum rate
\begin{eqnarray}
 R_{CC}^{\text{(sum)}}(W) & = & 4H(Z) \ = \ 4H(Y_1 + Y_3). \label{eq:ex_intro_temp2}
\end{eqnarray}

It will be shown in Theorem \ref{thm:chain_NCE} that there is a unique subspace chain decomposition of $V = <X_1, X_2, X_3,
X_4>$ given by $\{ \underline{0} \} \subsetneq W^{(1)} \subsetneq W^{(2)} \subsetneq W^{(3)} = V$, where
\begin{eqnarray}
 W^{(1)} & = & < X_1+X_2, X_2+X_3 > \nonumber \\
W^{(2)} & = & < X_1+X_2, X_2+X_3, X_3 + X_4 >. \label{eq:subpace_chain_intro}
\end{eqnarray}
With respect to this chain of subspaces, the best superspace $U \supseteq W$ to consider under the SS
approach is the smallest subspace in the chain that contains $W$, which in this case, is  $U = W^{(2)}$ (Fig.
\ref{fig:optimal_subspaces_Ex_intro}). The sum rate in this case, identified by Theorem
\ref{thm:chain_SSA}, turns out to be given by
\begin{eqnarray}
 R_{SS}^{\text{(sum)}}(W) & = & 4H(Y_3) \ < \ R_{CC}^{\text{(sum)}}(W), \label{eq:ex_intro_temp3}
\end{eqnarray}
where the second inequality in \eqref{eq:ex_intro_temp3} follows as the $\{Y_i\}$ are independent.
\begin{figure}[h!]
\begin{center}
\includegraphics[width=5cm]{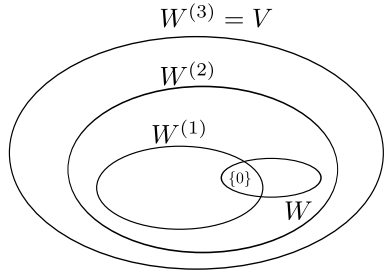}
\caption{The subspace chain decomposition of $V$ in Example  \ref{ex:opt_dist_intro} and its application in determining the
optimal subspace $U$ to compute $W$, under the SS approach. }
\label{fig:optimal_subspaces_Ex_intro} \end{center}
\end{figure}
\end{ex}

\vspace{0.1in}

\begin{figure}[h!]
\begin{center}
\includegraphics[width=6cm]{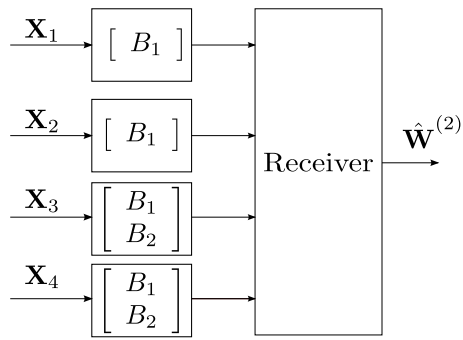}
\caption{Encoder structure under NC approach for decoding the subspace $W$ in
Example \ref{ex:opt_dist_intro}.}
\label{fig:nested_codes_intro} \end{center}
\end{figure}

\textit{Nested Codes (NC) approach} : This approach is motivated by the subspace-chain decomposition and may
be viewed as uniting under a common framework, both the CC approach of using a common linear encoder as well as the SW
approach of employing different encoders at each source. To illustrate the approach, we continue to work with Example
\ref{ex:opt_dist_intro}. Under the NC approach, the decoding happens in two stages. In the first stage the receiver, using
the CC approach decodes the subspace $W^{(1)}$. In the next stage, using $W^{(1)}$ as side information, $W^{(2)}$ is decoded
(using a modified CC approach which incorporates side information). The encoding matrices of the various sources are as shown
in Fig. \ref{fig:nested_codes_intro}. The matrix $B_1$ appearing in the figure is the common encoding matrix that would be
used if it was desired to compute subspace $W^{(1)}$ alone.  The block matrix $[B_1^T \ B_2^T]^T$ is the common encoder that
would have been used if one were only interested in computing the complement of $W^{(1)}$ in $W^{(2)}$ with $W^{(1)}$ as side
information.  It can be shown that there is a rearrangement of the $\{X_i\}$ under which the complement of $W^{(1)}$ in
$W^{(2)}$ can be made to be a function only of two of the random variables which, we have assumed without loss of generality
here to be $X_3,X_4$.  This explains why the submatrix $B_2$ appears only in the encoding of sources $X_3,X_4$.

The sum
rate in this case turns out to be given by
\bean
R^{\text{(sum)}}_{\text{NC}}(W) & = & 2H(Y_1)+2H(Y_3),
\eean
which can be shown to be less than $R_{SS}^{\text{(sum)}}(W)$ as well as the sum rate,
$R_{SW}^{\text{(sum)}}(W)=H(X_1,X_2,X_3,X_4)$, of the Slepian-Wolf (SW) approach under which the subspace $W$ is computed by
first recovering each of the four random variables $\{X_i\}$.

\begin{figure}[h!]
\begin{center}
\includegraphics[width=8cm]{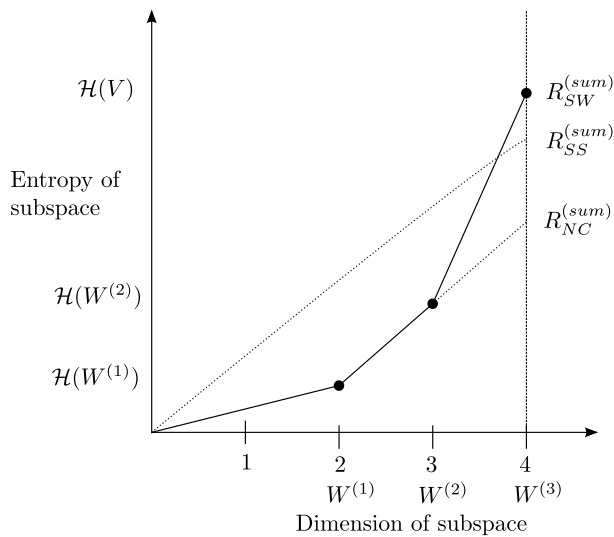}
\caption{Illustrating sum-rate calculation for the various approaches to computing $W$.}
\label{fig:graph_NC} \end{center}
\end{figure}

A graphical depiction of the sum rates achieved by the various schemes is provided in Fig.~\ref{fig:graph_NC}, with the sum rates appearing on the vertical axis on the far right.  It turns out that in general, we have
\bean
R^{\text{(sum)}}_{\text{CC}}(W) \ \geq \ R^{\text{(sum)}}_{\text{SS}}(W) \ \geq \ R^{\text{(sum)}}_{\text{NC}}(W), \\
R^{\text{(sum)}}_{\text{SW}}(W) \  \geq \ R^{\text{(sum)}}_{\text{NC}}(W).
\eean
In the particular case of the example, we have that there exist choices of probabilities $p_1,p_2$ such that
\bean
R^{\text{(sum)}}_{\text{SW}}(W) \ > \ R^{\text{(sum)}}_{\text{CC}}(W) \  > \ R^{\text{(sum)}}_{\text{SS}}(W) \ > \ R^{\text{(sum)}}_{\text{NC}}(W).
\eean

\subsection{Other Related Work}

Some early work on distributed function computation can be found in  \cite{HanKob}, \cite{AhlHan}, \cite{CsiKor}. In
\cite{HanKob}, the general problem of distributed function computation involving two sources is considered and the authors
identify conditions on the function such that the SW approach itself is optimal. In  \cite{AhlHan}, the authors address the
Korner-Marton problem of computing the modulo-two sum of two binary sources and produce an achievable-rate pair which is
strictly outside the time shared region of the SW and Korner-Marton rate regions\footnote{The sum rate of this achievable
rate-pair is however, still larger than the minimum of the SW and Korner-Marton sum rates.}.  For the same problem it is
shown in \cite{CsiKor}, that if $H(X_1 + X_2) \geq \min (H(X_1),H(X_2))$, then the SW scheme is sum-rate optimal. In
\cite{KriPra}, the authors showed how linear encoders are suitable for recovery of functions that are representable as
addition operations within an Abelian group.

The problem of compressing a source $X_1$, when $X_2$ is available as side information to the receiver, and where the
receiver is interested in decoding a function $f(X_1,X_2)$ under a distortion constraint, is studied in \cite{Yam}. For the
case of zero distortion, the minimum rate of compression is shown to be related to the conditional graph entropy of the
corresponding characteristic graph in \cite{Orl}. The extension of the nonzero distortion problem for the case of noisy
source and side information measurements is investigated in \cite{FenEffSav}.

In \cite{DosShaMedEff}, Doshi \textit{et. al.} consider the lossless computation of a function of two correlated,  but
distributed sources. They present a two-stage architecture, wherein in the first stage, the input sequence at each source is
divided into blocks of length $n$ and each block is coloured based on the corresponding characteristic graph at the source.
In the second stage, SW coding is used to compress the coloured data obtained at the output of first stage. The achievable
rate region using this scheme is given in terms of a multi-letter characterization and the optimality of the scheme is shown
for a certain class of distributions. In \cite{SefTch}, the authors derive inner and outer bounds for lossless compression of
two distributed sources $X,Y$ to recover a function $f(X,Y,Z)$, when $Z$ is available as side information to the receiver.
The bound is shown to be tight for partially invertible functions, i.e., for functions $f$ such that $X$ is a function of
$f(X,Y,Z)$ and $Z$.

For the case of two distributed Gaussian sources, computation of a linear combination of the sources is studied in
\cite{KritPra}, \cite{Wag}, wherein lattice-based schemes are shown to provide a rate advantage.  Zero-error function
computation in a network setting, has been investigated in \cite{KowKum},\cite{AppFraKarZeg},\cite{RaiDey}.

A notion of normalized entropy is introduced in Section \ref{sec:definitions}.   Section \ref{sec:CC_SS} discusses the
rate regions under the CC and SS approaches.  The unique decomposition of the $m$-dimensional space $V$ into a chain of
subspaces identified by a normalized measure of entropy, is presented in Section \ref{sec:struc_subspace}.  It is shown how
this simplifies determination of the minimum symmetric rate under the SS approach.  An example subspace computation along
with an attendant class of distributions for which the SS approach is optimal, are also presented here.      The nested-code
approach is presented in Section \ref{sec:nested_codes} along with conditions under which this approach improves upon the SW approach as well as examples for which the NC approach is sum-rate optimal. Most proofs are relegated to appendix.

\section{Normalized Entropy} \label{sec:definitions}

We will use $\rho_U$ to denote the dimension of a subspace $U$.

 {\textit{Entropy of a subspace}}: To every subspace $U$ of $V$, we will associate an entropy, which is the entropy of any set
of random variables that generate $U$. We will denote this quantity by $\mathcal{H}(U)$ and refer to this quantity loosely as
the \textit{entropy of the subspace}\footnote{We have used $\mathcal{H}(U)$ in place of $H(U)$ so as to avoid confusion with
the entropy of a random variable whose every realization is a subspace.} $U$.
Thus, if $U = <Y_1, \ldots, Y_{\rho_U}>$
\begin{eqnarray} \label{eq:entropy_subspace}
\mathcal{H}(U) & = & H(Y_1, \ldots, Y_{\rho_U}).
\end{eqnarray}
$\mathcal{H}(U)$ can also be viewed as the joint entropy of the collection of all random variables contained in the subspace
$U$, i.e., $\mathcal{H}(U) := H(\{U\})$.  Next, given any two subspaces $U_1$ and $U_2$, we define the conditional entropy of
the subspace $U_2$ conditioned on $U_1$ as
\begin{eqnarray}
\mathcal{H}(U_2|U_1) & \triangleq & H(\{U_2\}|\{U_1\}) \ .
\end{eqnarray} Let $U_1 + U_2$ denote the sum space of $U_1,U_2$.  Clearly, $\mathcal{H}(U_1 + U_2) =
H(\{U_1\},\{U_2\})$. Hence we can rewrite the above equation as
\begin{eqnarray} \label{eq:cond_entropy_subspace}
\mathcal{H}(U_2|U_1) & = & \mathcal{H}(U_1 + U_2) - \mathcal{H}(U_1).
\end{eqnarray}

\vspace{0.1in}

{\textit{Normalized entropy}}:
We define the {\em normalized entropy} $\mathcal{H}_N(U)$ of a non-zero subspace $U$ of $V$ as the entropy of $U$ normalized
by its dimension i.e.,
\begin{eqnarray} \label{eq:norm_entropy}
\mathcal{H}_N(U) & \triangleq & \frac{\mathcal{H}(U)}{\rho_U}.
\end{eqnarray}
For any pair of subspaces $U_1,U_2$, $ U_2 \nsubseteq U_1$, we define the {\em normalized, conditional entropy} of $U_2$
conditioned on $U_1$, to be given by
\begin{eqnarray} \label{eq:cond_norm_entropy}
\mathcal{H}_N(U_2|U_1) & \triangleq & \frac{\mathcal{H}(U_2|U_1)}{\rho_{U_2} - \rho_{U_1 \cap U_2}}.
\end{eqnarray}
Note that since $\rho_{U_1+U_2} \ = \ \rho_{U_1}+\rho_{U_2} - \rho_{U_1 \cap U_2}$, we can equivalently write
\begin{eqnarray} \label{eq:cond_norm_entropy_2}
\mathcal{H}_N(U_2|U_1) & \triangleq & \frac{\mathcal{H}(U_2+U_1) -\mathcal{H}(U_1)}{\rho_{U_2+U_1} - \rho_{U_1}}.
\end{eqnarray}

\begin{figure}[h!]
\begin{center}
\includegraphics[width=7cm]{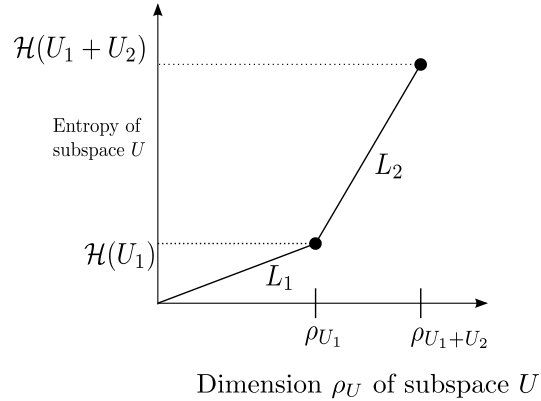}
\caption{An illustration of normalized entropies}
\label{fig:norm_cond_ent} \end{center}
\end{figure}

The above definitions are illustrated in Fig. \ref{fig:norm_cond_ent}, where the $x$-axis corresponds to the
dimension of subspaces and $y-$axis corresponds to the entropy of subspaces. The slope of the line $L_1$ is the normalized
entropy, ${\mathcal{H}}_N(U_1)$,  of $U_1$ and the slope of the line $L_2$ is the  normalized conditional entropy,
${\mathcal{H}}_N(U_2|U_1)$.

\section{Common Code and Selected Subspace Approaches} \label{sec:CC_SS}

The minimum symmetric rate of the CC and the SS approaches to the distributed subspace computation problem described in
Section \ref{sec:our_work}  are presented here.

\subsection{Rate Region Under the CC Approach} \label{sec:CC}

\begin{thm} \label{thm:CC}
Consider the distributed source coding setting shown in Fig. \ref{fig:CC_approach} where there are $m$ correlated sources $X_1, \ldots, X_m$ and receiver that is interested in decoding the $s$ dimensional subspace $W$ corresponding to the space spanned by the set $\{Z_i\}$ of random variables defined in \eqref{eq:prob_statement}. Then minimum symmetric rate under the CC approach is given by
\begin{eqnarray}
R_{\text{CC}}(W) & = & \max_{W_1 \subsetneq W} \mathcal{H}_N(W|W_1). \label{eq:CC_inf}
\end{eqnarray}
\end{thm}

\bpf
See Appendix \ref{app:CC_ach}.
\epf

\vspace*{.1in}

The best sum rate $R_{\text{CC}}^{\text{(sum)}}(W)$ under the CC approach is given by $R_{\text{CC}}^{\text{(sum)}}(W) =
mR_{\text{CC}}(W)$. Note that in the special case when the receiver is interested in just a single linear combination, $Z$,
of all the sources, the sum rate is simply $mH(Z)$. The following example illustrates the minimum symmetric rate for the
case, when the receiver is interested in decoding a two dimensional subspace.

\vspace*{.1in}

\begin{ex}
Let $m =3$ and the source alphabet be ${\mathbb{F}}_2$. Consider a receiver interested in computing $Z_1 = X_1, Z_2 =
X_2+X_3$ i.e, $W = <X_1, X_2+X_3>$. Then the minimum symmetric rate under the CC approach is given by
\bea
R_{\text{CC}}(W) & = & \max \{ \frac{H(Z_1,Z_2)}{2}, \ H(Z_1,Z_2|Z_1), \nonumber \\
&& \hspace*{.3in} H(Z_1,Z_2|Z_2), \ H(Z_1,Z_2|Z_1+Z_2)\}. \nonumber  \\
\eea
\end{ex}

\vspace{.05in}

\begin{note} \label{cor:CC_SW}
The CC approach is sum-rate optimal for the case when $W=V$ iff
\begin{eqnarray}
\mathcal{H}_N(V) & \geq & \mathcal{H}_N(V|V_1), \ \forall V_1 \subsetneq V.
\end{eqnarray}
This follows directly from Theorem \ref{thm:CC} by setting $W = V$ and noting that the optimal sum-rate in
this case is simply $H(X_1,\cdots,X_m)$.
\end{note}

\vspace{.1in}

\subsection{Rate Region Under the SS Approach} \label{sec:SS}

This approach recognizes that it is often more efficient to compute a superspace of $W$ rather than $W$ itself.  The identification of  the particular superspace that offers the greatest savings in compression rate is taken up in Section \ref{sec:struc_subspace}.

\vspace{.1in}

\begin{thm} \label{thm:SS}
Under the same setting as in Theorem~\ref{thm:CC}, the minimum symmetric rate under the SS approach is given by
\begin{align}
R_{\text{SS}}(W) & = \min_{U \supseteq W}\max_{U_1 \subsetneq U} \mathcal{H}_N(U|U_1).
\end{align}
\end{thm}

\bpf
Follows directly from Theorem \ref{thm:CC}.
\epf

\vspace{0.1in}

The (best) sum rate $R_{\text{SS}}^{\text{(sum)}}(W)$ under the SS approach is given by $R_{\text{SS}}^{\text{(sum)}}(W) =
mR_{\text{SS}}(W)$. Any subspace $U \supseteq W$ which minimizes $\max_{U_1 \subsetneq U} \mathcal{H}_N(U|U_1)$ will be
referred to as an \textit{optimal subspace} for computing $W$ under the SS approach.  There can be more than one optimal
subspace associated with a given $W$.

\section{A Decomposition Theorem for the Vector Space $V$ Based on Normalized Entropy} \label{sec:struc_subspace}

While the results of this section are used to identify the superspace $U$ that minimizes the quantity $\max_{U_1 \subsetneq U} \mathcal{H}_N(U|U_1)$ appearing in Theorem \ref{thm:SS}, they are also of independent interest as they exhibit an interesting interplay between linear algebra and probability theory.  Also included in this section, are example subspace-computation problems and a class of distributions for which the SS approach is sum-rate optimal, while the CC and the SW approaches are not.

\vspace*{0.1in}

\begin{thm} [Normalized-Entropy Subspace Chain] \label{thm:chain_NCE}
In the vector space $V$, there exists for some $r \leq m$, a unique, strictly increasing sequence of subspaces $\{ \underline{0} \} = W^{(0)} \subsetneq W^{(1)} \subsetneq  \ldots \subsetneq W^{(r)} = V $, such that, $\forall j \in \{1, \ldots, r \}$,
\begin{enumerate}
\item amongst all the subspaces of $V$ that strictly contain $W^{(j-1)}$, $W^{(j)}$ has the least possible value of normalized conditional entropy conditioned on $W^{(j-1)}$ and
\item if any other subspace that strictly contains $W^{(j-1)}$ also has the least value of normalized conditional entropy conditioned on $W^{(j-1)}$, then that subspace is strictly contained in $W^{(j)}$.
\end{enumerate}
Furthermore,
\begin{eqnarray}
\mathcal{H}_N(W^{(1)}|W^{(0)}) \ < \ \mathcal{H}_N(W^{(2)}|W^{(1)}) && \nonumber \\
&&\hspace*{-1.5in} < \ \ldots \  < \ \mathcal{H}_N(W^{(r)}|W^{(r-1)}) \label{eq:theorem_NCE_rates}.
\end{eqnarray}
\end{thm}
\bpf See Appendix \ref{app:proof_thm_chain_NCE}.
\epf

\vspace*{0.1in}

We illustrate Theorem \ref{thm:chain_NCE} below, by identifying the chain of subspaces $\{W^{(j)}\}$  for the case when the random variables $X_1, \ldots, X_m$ are derived via an invertible linear transformation of a set of $m$ statistically independent random variables $Y_1, \ldots, Y_m$.

\vspace{0.1in}

\begin{lem} \label{lem:ss_ind_sources}
Let $[X_1, \ldots, X_m] = [Y_1, \ldots, Y_m]G$, where $G$ is an $(m \times m)$ invertible matrix over $\mathbb{F}_q$ and $\{Y_i, i = 1, \ldots, m \}$ are $m$ independent random variables, each of which takes values in the finite field $\mathbb{F}_q$. Without loss of generality, let the entropies of $\{Y_i, i = 1, \ldots, m\}$ be ordered according to
\begin{align}
0 < H(Y_1) = \ldots = H(Y_{\ell_1}) < H(Y_{\ell_1+1}) = \ldots = \nonumber \\
&&&\hspace{-3in} H(Y_{\ell_1+ \ell_2}) < \ldots < H(Y_{\sum_{i=1}^{r-1}\ell_i+1}) = \ldots = H(Y_{\sum_{i=1}^{r}\ell_i}), \label{eq:entropy_order}
\end{align}
where $1 \leq \ell_i \leq m, i = 1, \ldots, r$ and $\sum_{i=1}^{r}\ell_i = m$. Then, the unique subspace chain identified by Theorem \ref{thm:chain_NCE} is given by
\begin{eqnarray} \label{eq:subspace_chain_indep}
\{\underline{0}\} \ \subsetneq \ <Y_1,\ldots,Y_{\ell_1}> \ \subsetneq \ <Y_1,\ldots, Y_{\ell_1+ \ell_2}> && \nonumber\\
&&\hspace*{-1.8in} \ \subsetneq \ldots  \subsetneq \ <Y_1,\ldots,Y_m> \ .
\end{eqnarray}
\end{lem}
\bpf See Appendix \ref{app:ind_dist_subspace_chain}.
\epf

\vspace*{.1in}

\begin{note}
While Theorem \ref{thm:chain_NCE} guarantees the existence of $r$, the above lemma shows that $r$ can take any value between
$1$ and $m$ depending on the joint distribution of the $\{X_i, 1 \leq i \leq m\}$.
\end{note}

\subsection{Identifying the Optimal Subspace Under the SS Approach}\label{sec:SS_simplified}

\begin{thm} [Optimal Rate under SS approach] \label{thm:chain_SSA}
Consider the distributed source coding problem shown in Fig. \ref{fig:CC_approach} having $m$ sources $X_1, \ldots, X_m$ and
a receiver that is interested in decoding the $s$ dimensional subspace $W$.  Let $W^{(0)} \subsetneq W^{(1)} \subsetneq
\ldots \subsetneq W^{(r)}$ be the unique subspace-chain decomposition of the vector space $V = < X_1, \ldots, X_m >$,
identified in Theorem \ref{thm:chain_NCE}.  Then an optimal subspace for decoding $W$ under the SS approach is given by
$U = W^{(j_0)}$, where $j_0$ is the unique integer, $1 \leq j_0 \leq r$, satisfying
\bean
W \ \subseteq \ W^{(j_0)}, & & W \ \nsubseteq \ W^{(j_0-1)} .
\eean
Furthermore,
\begin{eqnarray} \label{eq:optimal_rate_SS}
R_{\text{SS}}(W) \ = \ \mathcal{H}_N(W^{(j_0)}|W^{(j_0-1)}).
\end{eqnarray}
\bpf
See Appendix \ref{app:proof_thm_chain_SSA}.
\epf

\end{thm}

\vspace*{0.1in}

\begin{cor} \label{cor:chain_SSA}
With the $W^{(j)}, \ 1 \leq j \leq r$, as above,
\bean R_{\text{SS}}(W^{(j)}) \ = \ \mathcal{H}_N(W^{(j)}|W^{(j-1)}), & & \\
R_{\text{SS}}(W^{(1)}) < R_{\text{SS}}(W^{(2)}) < \ldots < R_{\text{SS}}(W^{(r)}) .
& & \eean
\end{cor}

\subsection{A subspace computation problem for which SS approach is sum-rate optimal} \label{sec:opt_dist}
Consider the setting where there are $m$ sources $X_1, \ldots X_m$ having a common alphabet $\mathbb{F}_2$, with $m$ even,
and a receiver interested in computing the sum $Z = (X_1 + \ldots + X_m) \mod 2$.  Let the joint distribution of
the $\{X_i, 1 \leq i \leq m\}$ be specified as follows:
\begin{equation} \label{eq:opt_dist_SS}
\left[\begin{array}{c} X_1 \\ X_2 \\ \vdots \\ X_{m-1} \\ X_m \end{array}\right] \ = \
\left[ \begin{array}{ccccc} 1 & 0 & \hdots  &  & 0\\
                     1 & 1 & 0 &  \hdots& 0 \\
                       &   &  \ddots & & \\
                      1 & 1  & \hdots     & 1 & 0\\
                      1 & 1  &  \hdots       & 1 & 1
                       \end{array}\right] \left[\begin{array}{c} Y_1 \\ Y_2 \\ \vdots \\ Y_{m-1} \\ Y_m \end{array}\right],
\end{equation}
where the $\{Y_i\}_{i=1}^{m}$ are statistically independent random variables such that $Y_i \sim
\text{Bernoulli}(\frac{1}{2})$, for $i$ odd and $Y_i \sim \text{Bernoulli}(p), 0 < p < \frac{1}{2}$, for $i$ even. When
$m=2$, $X_1$ and $X_2$ can be verified to possess a \textit{doubly-symmetric} joint distribution (see \cite{KorMar}), and
this is precisely the class of distribution for which  Korner and Marton showed that a common linear encoder is sum-rate
optimal for the computation of the modulo-2 sum, $Z = X_1 + X_2$. We now assume $m >2$ in the  above setting and show that
the SS approach yields optimal sum rate while the CC or SW approach do not. Note that by optimal sum rate we mean that this
is best sum rate that can be achieved for the subspace computation problem, even if encoders other than linear encoders were
permitted in Fig. \ref{fig:prob_statement}.

From Lemma \ref{lem:ss_ind_sources}, we know that the unique subspace chain for $V = < X_1, \ldots, X_m >$ is given by $\{0\}
\ \subsetneq  \ W^{(1)}  \ \subsetneq  \ W^{(2)}$, where
\begin{eqnarray}
 W^{(1)} & = & < Y_2, Y_4, \ldots, Y_m >  \nonumber \\
& = & < X_1 + X_2, X_3+X_4, \ldots, X_{m-1} + X_m > \nonumber
\end{eqnarray}
and $W^{(2)} = V$. Clearly, the subspace of interest $W = <X_1 + X_2 + \ldots + X_m> \ \subseteq \ W^{(1)}$ and hence by
Theorem $\ref{thm:chain_SSA}$, $W^{(1)}$ is an optimal subspace to decode $W$, under the SS approach. The minimum symmetric
rate is given by
\begin{equation}
R_{\text{SS}}(W) \ = \ \mathcal{H}_N(W^{(1)}) \ = \ \frac{H(Y_2, Y_4, \ldots, Y_m)}{\left( \frac{m}{2}\right)} \ = \ h(p), \nonumber
\end{equation}
yielding a sum rate $R_{\text{SS}}^{\text{(sum)}}  = mh(p)$.

Now, under the CC approach, since we directly decode the single linear combination $Z$, the sum rate is given by (Theorem
\ref{thm:CC})
\begin{eqnarray}
 R_{\text{CC}}^{\text{(sum)}} & = & m H(X_1+\ldots+X_m) \nonumber \\
& \stackrel{(a)}{=} & mH(Y_2+Y_4+\ldots+X_m) \ \stackrel{(b)}{>} \ R_{\text{SS}}^{\text{(sum)}}\nonumber,
\end{eqnarray}
where $(a)$ follows from \eqref{eq:opt_dist_SS} and $(b)$ follows since $\{Y_i\}$ are independent and $p < \frac{1}{2}$. .
Also, under the SW approach in which the whole space $V$ is first decoded before computing $W$, the sum rate is given by
$R_{\text{SW}}^{\text{(sum)}} \ = \ H(Y_1, \ldots, Y_m) \ = \ \frac{m}{2}(1+h(p)) > R_{\text{SS}}^{\text{(sum)}}$.

We now show that the SS approach is sum-rate optimal. If $(R_1, \ldots, R_m)$ is any achievable rate tuple, then $\forall i = 1, \ldots, m$, it must be true that
\begin{eqnarray}
R_{i} & \stackrel{(a)}{\geq} & H\left(X_1 + \ldots +X_m \left| X_1,\ldots,X_{i-1}, X_{i+1}, \ldots X_m \right. \right)  \nonumber \\
      &   =  & H(X_i | X_1,\ldots,X_{i-1}, X_{i+1}, \ldots X_m) \nonumber \\
      &   \stackrel{(b)}{=} & H(Y_1+\ldots+Y_i|Y_1, \ldots, Y_{i-1}, X_{i+1}, \ldots X_m)  \nonumber\\
      &   \stackrel{(c)}{=} & H(Y_i|Y_1, \ldots, Y_{i-1}, Y_i+Y_{i+1},Y_{i+2}, \ldots, Y_m) \nonumber \\
      &   \stackrel{(d)}{=}  & \left\{\begin{array}{cc}
      					H(Y_i|Y_i + Y_{i+1}), & i  \ <  \ m \\
      					H(Y_i),               & i \ =  \ m
        \end{array}\right. \nonumber \\
        & = & h(p), \label{eq:ex_conv_last_step}
\end{eqnarray}
where $(a)$ follows by considering a system in which we give $\{X_1,\ldots,X_m\} \backslash \{X_i\}$ as side information at
the receiver, $(b), (c)$ follow from \eqref{eq:opt_dist_SS} and $(d)$ follows from the independence of the $\{Y_i, i = 1,
\ldots, m\}$. The bound in \eqref{eq:ex_conv_last_step} holds true for all sources and hence $\sum_{i = 1}^{m}R_i \geq
mh(p)$. Since $R_{\text{SS}}^{\text{(sum)}} = mh(p)$, it follows that the SS approach is sum-rate optimal.

\section{Nested Codes Approach} \label{sec:nested_codes}

The NC approach to the subspace computation problem is a natural outgrowth of our decomposition theorem for the vector space $V$.  Under this approach, a sequential decoding procedure is adopted in which $W^{(j)}$ is decoded using $W^{(j-1)}$ as side information.

\subsection{CC Approach with Side Information} \label{sec:side_info}

As before,  we have a receiver that is interested in computing a subspace $W$ of $V$ with the difference this time, that the receiver possesses knowledge of a subspace $S$ of $W$,  $S = < Y_1, \ldots, Y_{\rho_S} >$ as side information. Let $T$ be a subspace of $W$ complementary to $S$ in $W$, i.e., $W$ is a direct sum of $S$ and $T$ which is denoted by $W = S \oplus T$. Then clearly, it suffices to compute $T$ given $S$ as side information.

We claim that there exists a complement $T$ of $S$ in $W$ which is a function of at most $(m - \rho_S)$ of the sources. This follows from noting that a basis for $S$ can be extended to a basis for $V$ by adding $(m - \rho_S)$ of the $\{X_i\}$.  Without loss of generality, we may assume that these are the random variables $X_{\rho_S+1}, \ldots, X_m$. Also, the intersection of $W$ with any complement of $S$ in $V$ is clearly a complement of $S$ in $W$. It follows that there is a complement $T$ of $S$ in $W$ which is only a function of the $(m - \rho_S)$ sources $X_{\rho_S+1}, \ldots, X_m$ and thus it is enough to encode the sources $X_{\rho_S+1}, \ldots, X_m$.

\begin{figure}[h!]
\begin{center}
\includegraphics[width=8cm]{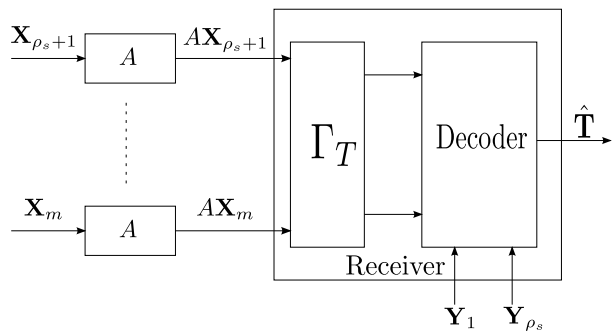}
\caption{CC approach for decoding with side information, when side information is linearly independent of the sources.}
\label{fig:CC_approach_with_side_info} \end{center}
\end{figure}

We adopt the CC approach here and hence, $n$-length realizations of all the
$(m - \rho_S)$ sources $X_{\rho_S+1}, \ldots, X_m$ are encoded by a common matrix encoder $A$. Now, if $T = < [X_{\rho_S+1}, \ldots, X_m]\Gamma_T>$ for some $(m-\rho_S) \times \rho_T$ matrix $\Gamma_T$ of rank $\rho_T$, then the receiver, as a first step, multiplies the received matrix $[A\bold{X}_{\rho_S+ 1} \ \ldots \ A\bold{X}_m]$ by $\Gamma_{T}$ on the right. These are then decoded using the side information $\bold{Y}_1, \ldots, \bold{Y}_{\rho_S}$ to obtain estimates of ${\bold{T}}$ (see Fig. \ref{fig:CC_approach_with_side_info}). The minimum symmetric rate for this approach is presented below.

\vspace*{.1in}

\begin{thm} \label{thm:CC_side_info}
Consider a distributed source coding problem, where the receiver is interested in computing the subspace $W$, given that $S \subsetneq W$ is available as side information to the receiver. The minimum symmetric rate under the CC based approach presented above, is given by
\begin{eqnarray}
R_{\text{CC}}(W|S) & = & \max_{T_1 \subsetneq T } \left\{\mathcal{H}_N(T|T_1 \oplus S) \right\} \label{eq:CC_side_info1}\\
& = & \max_{\substack{ W_1 \subsetneq W \\ s.t. W_1 \supseteq S}} \left\{\mathcal{H}_N(W|W_1) \right\} \label{eq:CC_side_info2}.
\end{eqnarray}
\end{thm}
\bpf  Similar to the proof of Theorem \ref{thm:CC}.
\epf

\vspace{0.1in}

Note from \eqref{eq:CC_side_info2} that irrespective of the particular complementary subspace $T$ that we choose to compute, the symmetric rate remains the same.  The specific choice of $T$ determines however, the number of $\{X_i, i = 1 \ldots m\}$ that are actually encoded.  Since $T$ has been selected such that
only $(m-\rho_S)$ sources are encoded, the achievable sum rate in this case, is given by $R_{\text{CC}}^{\text{(sum)}}(W|S) = (m-\rho_S)R_{\text{CC}}(W|S)$.

\vspace{0.1in}

\begin{cor} \label{cor:CC_side_info}
Consider the case where $W = W^{(j)}$ and $S=W^{(j-1)}$. Then
\begin{eqnarray}
R_{\text{CC}}(W^{(j)}|W^{(j-1)}) & = & \max_{\substack{ U_1 \subsetneq W^{(j)} \\ s.t. U_1 \supseteq W^{(j-1)}}} \left\{\mathcal{H}_N(W^{(j)}|U_1) \right\} \nonumber \\
& = & \mathcal{H}_N(W^{(j)}|W^{(j-1)}),
\end{eqnarray}
where the last equality follows since
\begin{eqnarray}
\mathcal{H}_N(W^{(j)}|W^{(j-1)}) & \leq & \max_{\substack{ U_1 \subsetneq W^{(j)} \\ s.t. U_1 \supseteq W^{(j-1)}}} \left\{\mathcal{H}_N(W^{(j)}|U_1) \right\} \nonumber \\
& \leq & \max_{\substack{ U_1 \subsetneq W^{(j)} }} \left\{\mathcal{H}_N(W^{(j)}|U_1) \right\} \nonumber \\
& \stackrel{(a)}{=} & \mathcal{H}_N(W^{(j)}|W^{(j-1)}),
\end{eqnarray}
where $(a)$ follows from Corollary \ref{cor:chain_SSA}. Thus,  $R_{\text{CC}}(W^{(j)}|W^{(j-1)}) = R_{\text{CC}}(W^{(j)})$, i.e., the rates per encoder are the same in this instance with and without side information. The difference between the two cases is that  in the presence of side information, we need encode only $(m- \rho_{W^{(j-1)}})$ sources as opposed to $m$ leading to a reduced sum rate by the fraction $\frac{(m- \rho_{W^{(j-1)}})}{m}$.   \end{cor}

\subsection{NC Approach for Subspace Computation} \label{subsec:nested_codes}
Consider the chain of subspaces $W^{(1)} \subsetneq \ldots \subsetneq
W^{(r)}$ as obtained from Theorem \ref{thm:chain_NCE}. Assume that we are interested in decoding the subspace $W^{(j)}, j
\leq r$. We will now describe a scheme for decoding $W^{(j)}$, that operates in $j$ stages. At stage $\ell, \ell \leq j$, we
decode $W^{(\ell)}$ using $W^{(\ell-1)}$ as side information, using the CC based approach described above in Section
\ref{sec:side_info}.  Using the same argument as in Section \ref{sec:side_info}, it follows that at stage $\ell, 1 \leq \ell
\leq j$, without loss of generality, it is enough to encode the sources $X_{m - \rho_{W^{(\ell-1)}}+1}, \ldots, X_m$.

From Corollary \ref{cor:CC_side_info}, the rate of each of the sources that are encoded in the $\ell^{th}$ stage is given by
\begin{equation}
R_{\ell} \triangleq R_{\text{CC}}(W^{(\ell)}|W^{(\ell-1)}) = \mathcal{H}_N(W^{(\ell)}|W^{(\ell-1)}).
\end{equation}
Also, let $A^{(\ell)}$ denote the common encoding matrix used in the $\ell^{th}$ stage. Since $R_1 < R_2 < \ldots < R_j$
(see Theorem \ref{thm:chain_NCE}), it can be shown, via a random coding argument and by invoking a union-bound argument on the probability of error calculation, that it is possible to choose the encoding matrices $A^{(1)}, \ldots, A^{(j)}$ having the following nested structure:
\begin{equation}
A^{(1)} = [B_1], A^{(2)} = \left[ \begin{array}{c} B_1 \\ B_2 \end{array} \right ] , \ldots, A^{(j)} = \left[
\begin{array}{c} B_1 \\ \vdots \\ B_j \end{array} \right ].
\end{equation}
Please see Appendix \ref{app:nested_codes} for a proof of this statement\footnote{Similar proofs regarding existence of nested linear codes have been shown in the past, for example see \cite{KriPra}.}.


Thus, the sum rate achieved for decoding the subspace $W^{(j)}$, under the NC approach, is given by
\begin{eqnarray}
\hspace*{-.1in}R_{\text{NC}}^{\text{(sum)}}(W^{(j)}) & & \nonumber  \\
&&\hspace*{-.9in}=\sum_{\ell=1}^{j-1} \left( \rho_{W^{(\ell)}} - \rho_{W^{(\ell-1)}} \right) \mathcal{H}_N(W^{(\ell)}|W^{(\ell-1)}) \nonumber \\
&&\hspace*{-.4in}+ (m - \rho_{W^{(j-1)}})\mathcal{H}_N(W^{(j)}|W^{(j-1)}) \nonumber \\
&& \hspace*{-.9in} = \mathcal{H}(W^{(j-1)}) + (m - \rho_{W^{(j-1)}})\mathcal{H}_N(W^{(j)}|W^{(j-1)}) \label{eq:sum_rate_nested_codes}.
\end{eqnarray}
As in the case of the SS approach, a scheme for decoding an arbitrary subspace $W$ under the NC approach would be to decode the
subspace $W^{(j_0)}$, where $j_0$ is the unique integer such that $W \subseteq W^{(j_0)}$ and $W \subsetneq W^{(j_0-1)}$.

Note that whereas the one-stage CC approach for decoding $W^{(j)}$ would have used the highest-rate matrix $A^{(j)}$ for all the sources, the NC approach uses it only for sources $X_{\rho_{W^{(j-1)}}+1},\ldots,X_m$ and uses lower-rate matrices for the remaining sources. Thus the NC approach clearly outperforms the SS approach for all subspaces with the exception of $W^{(1)}$. Even beyond this, the NC approach sum rate improves upon the SW sum rate for all subspaces $W \subseteq W^{(r-1)}$, while in all other cases it equals the SW sum rate. These comparisons are made explicit in the two theorems below.

\vspace{0.1in}

\begin{thm} \label{thm:nested_codes_comparison_SS}
The sum rate $R_{\text{NC}}^{\text{(sum)}}(W^{(j)})$ incurred in using the nested code approach for decoding the subspace $W^{(j)}, 1 \leq j \leq r$ satisfies $R_{\text{NC}}^{\text{(sum)}}(W^{(j)}) \leq mR_{\text{SS}}(W^{(j)})$, the sum rate for decoding $W^{(j)}$ using the SS approach. Equality holds iff $j=1$.
\end{thm}
\bpf
\begin{eqnarray}
R_{\text{NC}}^{\text{\text{sum}}}(W^{(j)}) \nonumber \\
&& \hspace*{-.5in}=\sum_{\ell=1}^{j-1} \left (\rho_{W^{(\ell)}} - \rho_{W^{(\ell-1)}}\right) \mathcal{H}_N(W^{(\ell)}|W^{(\ell-1)}) \nonumber \\
&&\hspace*{-.2in}+ (m-\rho_{W^{(j-1)}})\mathcal{H}_N(W^{(j)}|W^{(j-1)}) \nonumber \\
& &\hspace*{-.5in} \stackrel{(a)}{\leq} \sum_{\ell=1}^{j-1} \left(\rho_{W^{(\ell)}}- \rho_{W^{(\ell-1)}}\right)\mathcal{H}_N(W^{(j)}|W^{(j-1)}) \nonumber \\
&&\hspace*{-.2in} + (m-\rho_{W^{(j-1)}})\mathcal{H}_N(W^{(j)}|W^{(j-1)}) \nonumber \\
&& \hspace*{-.5in}=m \mathcal{H}_N(W^{(j)}|W^{(j-1)}) = mR_{\text{SS}}(W^{(j)}),
\end{eqnarray}
where $(a)$ follows by Theorem \ref{thm:chain_SSA}. Since the rates given in Theorem \ref{thm:chain_SSA} are strictly increasing, $(a)$ is an equality iff $j=1$.
\epf

\vspace*{0.1in}

\begin{thm} \label{thm:nested_codes_comparison_SW}
The sum rate $R_{\text{NC}}^{\text{(sum)}}(W^{(j)})$ incurred in using the nested code approach for decoding the subspace $W^{(j)}, 1 \leq j \leq r$ satisfies $R_{\text{NC}}^{\text{(sum)}}(W^{(j)}) \leq \mathcal{H}(V)$, the sum rate for decoding $W^{(j)}$ using the SW approach. Equality occurs iff $j=r$.
\end{thm}
\bpf
\begin{eqnarray}
\mathcal{H}(V) && \nonumber \\
&&\hspace*{-.5in}=\mathcal{H}(W^{(j-1)}) + \sum_{\ell=j}^{r}\mathcal{H}(W^{(\ell)}|W^{(\ell-1)}) \nonumber \\
&&\hspace*{-.5in}= \mathcal{H}(W^{(j-1)}) + \sum_{\ell=j}^{r} \left (\rho_{W^{(\ell)}} - \rho_{W^{(\ell-1)}}\right) \mathcal{H}_N(W^{(\ell)}|W^{(\ell-1)}) \nonumber \\
& &\hspace*{-.5in}\stackrel{(a)}{\geq}  \mathcal{H}(W^{(j-1)}) + \sum_{\ell=j}^{r} \left (\rho_{W^{(\ell)}} - \rho_{W^{(\ell-1)}}\right) \mathcal{H}_N(W^{(j)}|W^{(j-1)}) \nonumber \\
&& \hspace*{-.5in}=\mathcal{H}(W^{(j-1)}) + (m-\rho_{W^{(j-1)}})\mathcal{H}_N(W^{(j)}|W^{(j-1)}) \nonumber \\
&& \hspace*{-.5in}=R_{\text{NC}}^{\text{\text{sum}}}(W^{(j)}) \nonumber,
\end{eqnarray}
where $(a)$ follows from Theorem \ref{thm:chain_SSA}. Since the rates given in Theorem \ref{thm:chain_SSA} are strictly increasing, $(a)$ holds with equality iff $j=r$.
\epf

\subsection{An example subspace computation problem for which NC approach is optimal} \label{sec:opt_dist2}
We now revisit Example \ref{ex:opt_dist_intro} introduced in Section \ref{sec:intro} and show that the NC approach is
sum-rate optimal if the subspace of interest is $W = W^{(2)}$. It is not hard to show that the subspace chain decomposition
for the joint distribution in Example \ref{ex:opt_dist_intro} is indeed as given in \eqref{eq:subpace_chain_intro}. Thus,
from \eqref{eq:sum_rate_nested_codes}, the sum rate achievable using the NC scheme is given by
\begin{eqnarray} \label{eq:achievable_rate_nested_codes_ex}
R_{\text{NC}}^{\text{sum}}(W^{(2)}) & = & \mathcal{H}(W^{(1)}) + (m - \rho_{W^{(1)}})\mathcal{H}_N(W^{(2)}|W^{(1)}) \nonumber \\
 & = & 2h(p_1) + 2h(p_2).
\end{eqnarray}
To show sum-rate optimality, note that if $(R_1, R_2, R_3, R_4)$ is
any achievable rate tuple, then we must have
\begin{eqnarray}
R_4 & \stackrel{(a)}{\geq} & H(X_1+X_2, X_2+X_3, X_3+X_4|X_1,X_2,X_3) \nonumber \\
& = & H(X_4|X_1,X_2,X_3) \nonumber \\
& \stackrel{(b)}{=} & H(Y_4|Y_1,Y_2,Y_3+Y_4) \nonumber \\
& \stackrel{(c)}{=} & H(Y_4|Y_3+Y_4) = h(p_2)\label{eq:outerbound_1_nested_codes_ex},
\end{eqnarray}
where $(a)$ follows by a considering a system in which $X_1, X_2, X_3$ is given as side information, $(b)$ follows from \eqref{eq:ex_intro_temp1} and
$(c)$ follows from the independence of $\{Y_i, i = 1, \ldots, 4\}$. Next, if we consider a second system in which $X_4$ alone is
given as side information, then it must be that
\begin{eqnarray}
R_1 + R_2 + R_3 & \geq & H(X_1+X_2, X_2+X_3, X_3+X_4|X_4) \nonumber \\
& = & H(Y_1,Y_2,Y_3|Y_4) \nonumber \\
& = & h(p_2) + 2h(p_1)\label{eq:outerbound_2_nested_codes_ex}.
\end{eqnarray}
Combining \eqref{eq:outerbound_1_nested_codes_ex} and \eqref{eq:outerbound_2_nested_codes_ex}, we get the lower bound on the sum rate given by $R_1 + R_2 + R_3 + R_4 \geq 2h(p_1) + 2h(p_2)$. This, along with \eqref{eq:achievable_rate_nested_codes_ex} implies sum-rate optimality of
the NC approach. Note from Theorems \ref{thm:nested_codes_comparison_SS} and \ref{thm:nested_codes_comparison_SW}
that the subspace and the SW approaches are both strictly suboptimal in this case.

\bibliographystyle{IEEEtran}
\bibliography{JSAC_bib}

\begin{thebibliography}{10}
\providecommand{\url}[1]{#1}
\csname url@samestyle\endcsname
\providecommand{\newblock}{\relax}
\providecommand{\bibinfo}[2]{#2}
\providecommand{\BIBentrySTDinterwordspacing}{\spaceskip=0pt\relax}
\providecommand{\BIBentryALTinterwordstretchfactor}{4}
\providecommand{\BIBentryALTinterwordspacing}{\spaceskip=\fontdimen2\font plus
\BIBentryALTinterwordstretchfactor\fontdimen3\font minus
  \fontdimen4\font\relax}
\providecommand{\BIBforeignlanguage}[2]{{%
\expandafter\ifx\csname l@#1\endcsname\relax
\typeout{** WARNING: IEEEtran.bst: No hyphenation pattern has been}%
\typeout{** loaded for the language `#1'. Using the pattern for}%
\typeout{** the default language instead.}%
\else
\language=\csname l@#1\endcsname
\fi
#2}}
\providecommand{\BIBdecl}{\relax}
\BIBdecl

\bibitem{LalPraVinKumPra}
V.~Lalitha, N.~Prakash, K.~Vinodh, P.~V. Kumar, and S.~S. Pradhan, ``A nested
  linear codes approach to distributed function computation over subspaces,''
  in \emph{Proc. Allerton Conf. Communication, Control and Computing},
  Monticello, IL, Sep. 2011.

\bibitem{KorMar}
J.~Korner and K.~Marton, ``How to encode the modulo-two sum of binary sources
  (corresp.),'' \emph{IEEE Trans. Inf. Theory}, vol.~25, no.~2, pp. 219--221,
  Mar. 1979.

\bibitem{SleWol}
D.~Slepian and J.~Wolf, ``Noiseless coding of correlated information sources,''
  \emph{IEEE Trans. Inf. Theory}, vol.~19, no.~4, pp. 471--480, Jul. 1973.

\bibitem{HanKob}
T.~S. Han and K.~Kobayashi, ``A dichotomy of functions $f(x, y)$ of correlated
  sources $(x, y)$ from the viewpoint of the achievable rate region,''
  \emph{IEEE Trans. Inf. Theory}, vol.~33, no.~1, pp. 69 -- 76, Jan. 1987.

\bibitem{AhlHan}
R.~Ahlswede and T.~S. Han, ``On source coding with side information via a
  multiple-access channel and related problems in multi-user information
  theory,'' \emph{IEEE Trans. Inf. Theory}, vol.~29, no.~3, pp. 396 -- 412, May
  1983.

\bibitem{CsiKor}
I.~Csiszar and J.~Korner, \emph{Information theory: coding theorems for
  discrete memoryless systems}.\hskip 1em plus 0.5em minus 0.4em\relax New
  York: Academic, 1981.

\bibitem{KriPra}
D.~Krithivasan and S.~S. Pradhan, ``Distributed source coding using abelian
  group codes: A new achievable rate-distortion region,'' \emph{IEEE Trans.
  Inf. Theory}, vol.~57, no.~3, pp. 1495 --1519, Mar. 2011.

\bibitem{Yam}
H.~Yamamoto, ``Wyner - ziv theory for a general function of the correlated
  sources (corresp.),'' \emph{IEEE Trans. Inf. Theory}, vol.~28, no.~5, pp. 803
  -- 807, Sep. 1982.

\bibitem{Orl}
A.~Orlitsky and J.~R. Roche, ``{Coding for computing},'' \emph{IEEE Trans. Inf.
  Theory}, vol.~47, no.~3, pp. 903--917, Mar. 2001.

\bibitem{FenEffSav}
H.~Feng, M.~Effros, and S.~Savari, ``Functional source coding for networks with
  receiver side information,'' in \emph{Proc. Allerton Conf. Communication,
  Control and Computing}, Sep. 2004, pp. 1419--1427.

\bibitem{DosShaMedEff}
V.~Doshi, D.~Shah, M.~Medard, and M.~Effros, ``Functional compression through
  graph coloring,'' \emph{IEEE Trans. Inf. Theory}, vol.~56, no.~8, pp. 3901
  --3917, Aug. 2010.

\bibitem{SefTch}
M.~Sefidgaran and A.~Tchamkerten, ``Computing a function of correlated sources:
  A rate region,'' in \emph{Proc. IEEE Int. Symp. Information Theory}, St.
  Petersburg, 2011, pp. 1856 --1860.

\bibitem{KritPra}
D.~Krithivasan and S.~S. Pradhan, ``Lattices for distributed source coding:
  Jointly gaussian sources and reconstruction of a linear function,''
  \emph{IEEE Trans. Inf. Theory}, vol.~55, no.~12, pp. 5628 --5651, Dec. 2009.

\bibitem{Wag}
A.~B. Wagner, ``On distributed compression of linear functions,'' \emph{IEEE
  Trans. Inf. Theory}, vol.~57, no.~1, pp. 79 --94, Jan. 2011.

\bibitem{KowKum}
H.~Kowshik and P.~R. Kumar, ``Zero-error function computation in sensor
  networks,'' in \emph{Proc. IEEE Conf. Decision and Control}, 2009, pp.
  3787--3792.

\bibitem{AppFraKarZeg}
R.~Appuswamy, M.~Franceschetti, N.~Karamchandani, and K.~Zeger, ``Network
  coding for computing: Cut-set bounds,'' \emph{IEEE Trans. Inf. Theory},
  vol.~57, no.~2, pp. 1015 --1030, Feb. 2011.

\bibitem{RaiDey}
B.~K. Rai and B.~K. Dey, ``On network coding for sum-networks,'' \emph{IEEE
  Trans. Inf. Theory}, vol.~58, no.~1, pp. 50 --63, Jan. 2012.

\bibitem{Kra}
G.~Kramer, ``{Topics in multi-user information theory},'' \emph{Found. Trends
  Commun. Inf. Theory}, vol.~4, no. 4-5, pp. 265--444, 2007.

\bibitem{CovTho}
T.~M. Cover and J.~A. Thomas, \emph{{Elements of information theory}},
  2nd~ed.\hskip 1em plus 0.5em minus 0.4em\relax New York: Wiley, 2006.

\end{thebibliography}

\appendices

\section{Proof of Theorem \ref{thm:CC}} \label{app:CC_ach}

Before proceeding to prove the theorem, for notational simplicity we shall denote $[Z_1,\ldots Z_s]$ by $Z^{[1:s]}$ and assume that $W_1$ is generated by $Z^{[1:s]}G$, where $G$ is a full rank $s \times \nu$ matrix, where $\nu = \rho_{W_1}$. Thus, the set of achievable rates per encoder under the CC approach given by Theorem \ref{thm:CC} can be rewritten as follows.
\begin{equation}
{\mathcal R}_{\text{CC}}(W) \ =  \ \left \{ R \mid R \geq  \frac{1}{(s-\nu)} H(Z^{[1:s]} \mid Z^{[1:s]}G) \right\} \label{eq:rate_CC_eq_vec}
\end{equation}
for every choice of $G$, whose column space corresponds to a $\nu$-dimensional subspace of $\mathbb{F}_q^s$, $ 0 \leq \nu \leq s-1$.

\begin{figure}[h!]
\begin{center}
\includegraphics[width=8cm]{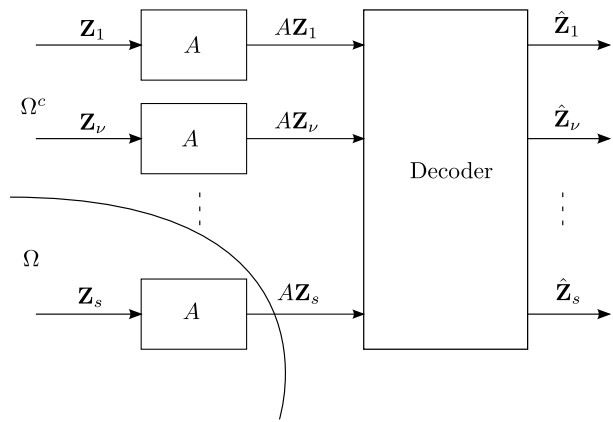}
\caption{Equivalent system model for the CC approach}
\label{fig:eq_sys_mod_km} \end{center}
\end{figure}

In order to prove the theorem, we shall work with the system model shown in Fig.~\ref{fig:eq_sys_mod_km}, which is equivalent to that of CC approach. This system is same as SW system except that all the sources are encoded by a common matrix $A$. A rate $R$ per encoder is achievable in this equivalent system iff it is achievable in the original system of interest (see Fig.~\ref{fig:CC_approach}).

{\em Achievability}: The notion of typical sets will be required to prove the achievability. The following definition of  $\epsilon-$typical set of a random variable $X$ will be used:
\begin{equation}
A_{\epsilon}^{(n)}(X)   =  \left \{ \bold{x}:
\left |\frac{N(a|\bold{x})}{n} - P_X(a) \right | \leq \epsilon P_X(a),  \forall a \in \mathcal{X} \right \} \ , \label{eq:typ_set_def}
\end{equation}
where $\mathcal{X}$ is the alphabet of the random variable $X$, $N(a|\bold{x})$ denotes the number of occurrences of the symbol $a$ in the realization $\bold{x}$.  We refer the reader to \cite{Kra} for properties of this typical set as well as the related notions of conditional and joint typical sets. We shall make use of the following lemma in the proof of the achievability.

\begin{lem} \label{lem:count_typ_seq}
Let $X$ and $Y$ be two discrete random variables taking on values over a finite alphabet $\mathcal{X}$ and $\mathcal{Y}$ respectively. Let $Y = f(X)$ be a deterministic function of $X$. Let ${\bf x}$ be an $n-$length realization of the i.i.d. random variable ${X}$ and ${\bf y} = f^n({\bf x})$, where $f^n({\bf y}) = (f(y_1), \ldots , f(y_n))$.
Then we have
\bea
    \left\{{\bf x} \in A_{\epsilon}^{(n)}(X) \mid f^n({\bf x}) = {\bf y} \right\} & = & A_{\epsilon}^{(n)}(X|{\bf y}). \label{eq:count_typ_seq}
\eea
\end{lem}

\bprf The proof can shown by using the definition of the typical set as given by \eqref{eq:typ_set_def}.
\eproof

Achievability will be shown using a random coding argument by averaging over set of all matrix encoders of the form $A : \mathbb{F}_q^n  \longrightarrow  \mathbb{F}_q^k$, where the $k \times n$ matrix $A$ is assumed to be a realization of the random matrix ${\bf A}$, distributed uniformly on the ensemble $M_{k \times n}(\mathbb{F}_q)$. We will apply the joint typical set decoder and calculate the probability of error $P_e^{(n)}$ averaged over the all the source symbols and also over all realizations of ${\bf A}$.

Let the source sequences to be ${\bf z}^{[1:s]}$. Then the decoder will declare ${\bf \hat{z}}^{[1:s]}$ to be the transmitted sequence if it is the unique sequence that belongs to $A_{\epsilon}^{(n)}(Z^{[1:s]})$ and $A{\bf \hat{z}}^{[1:s]} = A{\bf z}^{[1:s]}$.  Thus, the decoder will make an error if any one of the following events happen:
\begin{eqnarray}
    E_1 & : & {\bf z}^{[1:s]}  \notin  A_{\epsilon}^{(n)}(Z^{[1:s]}) \\
    E_2 & : & \exists {\bf v}^{[1:s]} \in A_{\epsilon}^{(n)}(Z^{[1:s]}) \ \text{such that} \nonumber \\
     && \ {\bf v}^{[1:s]} \neq {\bf z}^{[1:s]} \  \text{and}  \ A{\bf v}^{[1:s]} = A{\bf z}^{[1:s]}.
\end{eqnarray}
Let us denote ${\bf \Delta}^{[1:s]} = {\bf v}^{[1:s]} - {\bf z}^{[1:s]}$. Then, the probability of error is upper bounded as
\begin{eqnarray} \label{eq:fields_PE1_PE2}
&& \hspace*{-.4in} P_{e}^{(n)} \leq P(E_1) + P(E_2) \nonumber \\
&& \hspace*{-.1in} \leq \delta_n + \nonumber \\
&& \hspace*{-.2in} \sum_{{\bf z}^{[1:s]} \in A_{\epsilon}^{(n)}(Z^{[1:s]}) }P({\bf z}^{[1:s]})\underbrace{\sum_{\substack{{\bf v}^{[1:s]} \in A_{\epsilon}^{(n)}( Z^{[1:s]}) : \\    {\bf v}^{[1:s]} \neq {\bf z}^{[1:s]}}} P({\bf A} {\bf \Delta}^{[1:s]} = 0)}_{P_1} , \nonumber \\
\end{eqnarray}
where $\delta_n \rightarrow 0$ as $n \rightarrow \infty$.
We will now compute $P({\bf A} {\bf \Delta}^{[1:s]} = 0)$ as follows. Let ${\cal N}({\bf \Delta}^{[1:s]})$ be the nullspace of ${\bf \Delta}^{[1:s]}$ and $\nu$ be its rank. Since ${\bf \Delta}^{[1:s]} \neq 0$, we have $ 0 \leq \nu \leq s-1$. The rank of ${\bf \Delta}^{[1:s]} $ is $(s-\nu)$ and hence the rank of the left nullspace of ${\bf \Delta}^{[1:s]}$ is $n-(s-\nu)$. Thus, the number of matrices which satisfy ${\bf A} {\bf \Delta}^{[1:s]} = 0$ is $(q^{n-(s-\nu)})^k$. Since there are $q^{kn}$ choices for the matrix $A$, we get
\bea
    P({\bf A} {\bf \Delta}^{[1:s]} = 0) = \frac{(q^{n-(s-\nu)})^k}{q^{kn}} = q^{-k(s-\nu)} \label{eq:prob_delta}.
\eea
Thus partitioning the set of all ${\bf \Delta}^{[1:s]}$ based on the rank of ${\cal N}({\bf \Delta}^{[1:s]})$, we can rewrite $P_1$ in \eqref{eq:fields_PE1_PE2} as
\bea \label{p_e_field1}
    P_1 & \leq & \sum\limits_{\nu=0}^{s-1} \ \sum\limits_{W_1: \ \text{dim}(W_1)=\nu} \  \sum_{\substack{{\bf v}^{[1:s]} \in
A_{\epsilon}^{(n)}( Z^{[1:s]}):  \\    {\cal N}({\bf \Delta}^{[1:s]}) = W_1}} \hspace{-0.2in}q^{-k(s-\nu)},
\eea
where $W_1$ is a $\nu$ dimensional subspace of $\mathbb{F}_q^s$. We shall now provide an alternative expression for the set $\{{\bf v}^{[1:s]} \in A_{\epsilon}^n(Z^{[1:s]}) \ | \ {\cal N}({\bf \Delta}^{[1:s]}) = W_1 \}$ as follows. Let $\{ {\bf g}_1, {\bf g}_2, \ldots , {\bf g}_{\nu}\}$ denote a basis for $W_1$ and let $G_{W_1} = [{\bf g}_1  \ldots  {\bf g}_{\nu}]$. Then
\bea
    \left\{{\bf v}^{[1:s]} \in A_{\epsilon}^{(n)}(Z^{[1:s]}) \mid {\cal N}({\bf \Delta}^{[1:s]})= W_1 \right\} && \nonumber\\
    && \hspace*{-2.6in}= \left\{ {\bf v}^{[1:s]} \in A_{\epsilon}^{(n)}(Z^{[1:s]}) \mid {\bf \Delta}^{[1:s]} G_{W_1} = 0 \right\} \nonumber \\
    & & \hspace*{-2.6in}=\left\{ {\bf v}^{[1:s]} \in A_{\epsilon}^{(n)}(Z^{[1:s]}) \mid {\bf v}^{[1:s]} G_{W_1} = {\bf z}^{[1:s]} G_{W_1} \right\}.
\eea
Now applying Lemma \ref{lem:count_typ_seq} to the above equation wherein we set $X = Z^{[1:s]}$, $f(X) = Z^{[1:s]}G_{W_1}$ and by noting that ${\bf v}^{[1:s]}$ is an $n-$length realization of $Z^{[1:s]}$, we get
\bea
    \left\{{\bf v}^{[1:s]} \in A_{\epsilon}^{(n)}(Z^{[1:s]}) \mid {\bf v}^{[1:s]} G_{W_1} = {\bf z}^{[1:s]} G_{W_1} \right\} &&\nonumber \\
    & & \hspace*{-2in}= A_{\epsilon}^{(n)}(Z^{[1:s]} \mid {\bf z}^{[1:s]} G_{W_1}).
\eea
We substitute the above equation in \eqref{p_e_field1} and use the resulting expression in \eqref{eq:fields_PE1_PE2} to get
\bea
    P_{e}^{(n)} & \leq & \delta_n + \sum_{{\bf z}^{[1:s]} \in A_{\epsilon}^n(Z^{[1:s]})} P({\bf z}^{[1:s]})\sum\limits_{\nu=0}^{s-1} \ \nonumber \\
    &&\hspace*{-.5in}   \sum \limits_{W_1: \ \text{dim}(W_1)=\nu} 2^{n[H(Z^{[1:s]} \mid Z^{[1:s]}G_{W_1} )(1+\epsilon)-(s-\nu)\frac{k}{n} \log (q) ]}.  \nonumber \\   \label{p_e_field2}
\eea
where we used the fact that the size of the conditional typical set is bounded as
\begin{equation}
\left | A_{\epsilon}^{(n)}(Z^{[1:s]} \mid {\bf z}^{[1:s]} G_{W_1}) \right | \leq 2^{nH(Z^{[1:s]} \mid Z^{[1:s]}G_{W_1} )(1+\epsilon)}.
\end{equation}
Thus, a sufficient condition for $P_e^{(n)}\rightarrow 0$ is that
\bea
    \frac{k}{n} \log (q) > \frac{1}{(s-\nu)}H(Z^{[1:s]} \mid Z^{[1:s]}G_{W_1} )(1+\epsilon)
\eea
for every choice of $\nu$-dimensional subspace $W_1$ of $\mathbb{F}_q^s$, $ 0 \leq \nu \leq s-1$.

We will now show the necessity of the inequalities in \eqref{eq:rate_CC_eq_vec} if reliable decoding of the sources $Z^{[1:s]}$ is desired thereby proving that $R_{\text{CC}}(W)$ is the minimum symmetric rate achievable under CC approach. Let $(\Omega, \Omega^c)$ denote a partition of the sources $Z^{[1:s]}$, such that $|\Omega| = (s- \nu), 0 \leq \nu \leq (s-1)$ and $Z^{\Omega} = \{Z_j, j \in \Omega \}$. It follows from SW lower bound  \cite{CovTho} that
\bea
    R & \geq &  \frac{1}{(s-\nu)}H(Z^{\Omega} \mid Z^{\Omega^c}) \label{eq:cc_converse_sw}
\eea
is a necessary condition. Note that the above inequalities are exactly those  in \eqref{eq:rate_CC_eq_vec} obtained by choosing the columns of $G$ from the set of standard basis vectors for $\mathbb{F}_q^s$.

\begin{figure}[h]
\begin{center}
\includegraphics[width=8cm]{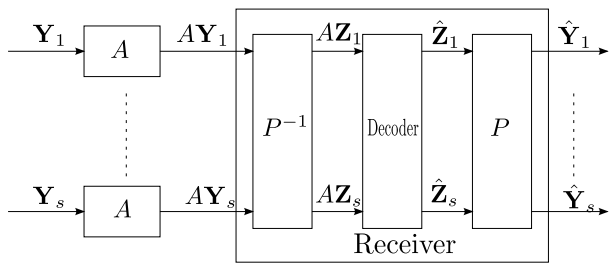}
\caption{A derived system to compute $Y^{[1:s]}$} \label{fig:cut_set_bound_fields2}
\end{center}
\end{figure}

We will now show that the necessity of remaining inequalities in \eqref{eq:rate_CC_eq_vec}, corresponding to other choices of  $G$, is due to our restriction to a common encoding matrix. Consider a new system (which is also reliable) constructed as shown in Fig. \ref{fig:cut_set_bound_fields2} with the same encoder and decoder as that of the system in Fig. \ref{fig:eq_sys_mod_km}, where $Y^{[1:s]} = Z^{[1:s]}P$, $P$ being an $s \times s$ invertible matrix.
Applying the SW bounds \cite{CovTho} to this new system we get,
\bea
    R & \geq  &   \frac{1}{(s-\nu)} H(Y^{\Omega} \mid Y^{\Omega^c})   \ ,
\eea
where $\Omega$ is some subset of the sources $Y^{[1:s]}$, such that $|\Omega| = (s- \nu), 0 \leq \nu \leq (s-1)$. Since, $Y^{[1:s]} = Z^{[1:s]}P$ and $P$ is an invertible matrix, the above equation can be written as
\bea
    R  & \geq & \frac{1}{(s-\nu)}H(Z^{[1:s]} \mid Z^{[1:s]}G)  \ ,
\eea
where $G$ is an $s \times \nu$ sub matrix of $P$ containing the $\nu$ columns corresponding to $Y^{\Omega^c}$. Since the above bound is true for every invertible matrix $P$, we run through all subspaces of $\mathbb{F}_q^s$ of dimension less than or equal to $s-1$ thereby establishing the necessity of the inequalities in \eqref{eq:rate_CC_eq_vec}.

\section{Proof of Theorem \ref{thm:chain_NCE}} \label{app:proof_thm_chain_NCE}

We will first present a few properties of normalized and conditional normalized entropies, which will subsequently be used to prove the theorem.

\vspace{0.1in}

\begin{lem} \label{lem:NE_1}
Consider $U, W \subseteq V$ such that $U \nsubseteq W$. Then
\begin{equation} \label{eq:NE_lem1_1}
\mathcal{H}_N(U+W_1|W) = \mathcal{H}_N(U|W), \ \ \forall \ W_1 \subseteq W.
\end{equation}
\end{lem}
\bprf Follows directly by invoking the equivalent definition of conditional normalized entropy in \eqref{eq:cond_norm_entropy_2}.

\vspace{0.1in}

\begin{lem} \label{lem:NE_3}
Consider $U, W \subseteq V$ such that $U \nsubseteq W$. Then
\begin{equation}
\mathcal{H}_N(U|W) \leq \mathcal{H}_N(U|U \cap W).
\end{equation}
\end{lem}
\bprf Follows from the definition of conditional normalized entropy in \eqref{eq:cond_norm_entropy} and by using that fact that $\mathcal{H}(U|W) \leq \mathcal{H}(U|U \cap W)$.

\vspace{0.1in}

\begin{lem} \label{lem:NE_2}
Consider $W, U_1, U \subseteq V$ such that $W \subsetneq U_1 \subsetneq U$. Then, one of the following three conditions is true.
\begin{eqnarray*}
\text{a) }\hspace{0.1in}\mathcal{H}_N(U_1|W) < & \mathcal{H}_N(U|W) & < \mathcal{H}_N(U|U_1) \label{eq:NE2_lt}. \\
\text{b) }\hspace{0.1in}\mathcal{H}_N(U_1|W) = & \mathcal{H}_N(U|W) & = \mathcal{H}_N(U|U_1) \label{eq:NE2_eq}. \\
\text{c) }\hspace{0.1in}\mathcal{H}_N(U_1|W) > & \mathcal{H}_N(U|W) & > \mathcal{H}_N(U|U_1) \label{eq:NE2_gt}.
\end{eqnarray*}
\end{lem}
\bprf
$\mathcal{H}_N(U|W)$ can be written as a convex combination of $\mathcal{H}_N(U_1|W)$ and $\mathcal{H}_N(U|U_1)$ as
\begin{equation}
\mathcal{H}_N(U|W) = \alpha \mathcal{H}_N(U_1|W) + (1-\alpha) \mathcal{H}_N(U|U_1),
\end{equation}
where $\alpha = \frac{\rho_{U_1}-\rho_W}{\rho_U-\rho_W}$. The lemma now follows.
\eproof

\vspace{0.1in}

\noindent {\em Proof of Theorem \ref{thm:chain_NCE}}: Consider the set
\bea
S_{W_0} &= & \left\{ U |  W_0 \subsetneq U \ \text{and} \ \mathcal{H}_N(U|W_0) \leq \mathcal{H}_N(W|W_0) \right. \nonumber \\
&& \hspace*{.5in}\left. \forall \ W \subseteq V, W_0 \subsetneq W\right\},
\eea
i.e., $S_{W_0}$ is the set of all subspaces of $V$ which contain $W_0$ and have the least normalized conditional entropy conditioned on $W_0$.
We claim that $S_{W_0}$ is closed under subspace addition, i.e., if $U_1, U_2 \in S_{W_0}$, then $U_1+U_2 \in S_{W_0}$. The claim will be proved shortly. Since $S_{W_0}$ is a finite set, this will imply that $Q(W_0) \triangleq \sum_{U \in S_{W_0}} U$ is the unique maximal element of $S_{W_0}$. Now, consider the chain obtained sequentially as follows:
\begin{equation} \label{eq:chain_construciton}
W^{(j)} \triangleq Q\left(W^{(j-1)}\right), \ \forall \ j > 1,
\end{equation}
where $W^{(0)} = \{\underline{0}\}$.  The construction proceeds until the $r^{\text{th}}$ stage, where $ W^{(r)} = V$. It is clear that the chain obtained from \eqref{eq:chain_construciton} satisfies conditions $1)$ and $2)$ in Theorem \ref{thm:chain_NCE} and is also unique. To prove that this chain also satisfies \eqref{eq:theorem_NCE_rates}, apply Lemma \ref{lem:NE_2} to the three element subspace chain $W^{(j-1)} \subsetneq W^{(j)} \subsetneq W^{(j+1)}, \ 1 \leq j \leq r-1$. Since $W^{(j)} = Q(W^{(j-1)})$, we have that $\mathcal{H}_N(W^{(j)}|W^{(j-1)})  <  \mathcal{H}_N(W^{(j+1)}|W^{(j-1)})$. Hence condition $\text{a})$ of Lemma \ref{lem:NE_2} is true in this case and thus
\begin{equation}
\mathcal{H}_N(W^{(j)}|W^{(j-1)}) \  <  \ \mathcal{H}_N(W^{(j+1)}|W^{(j)}).
\end{equation}

Now, we will prove our claim that $S_{W_0}$ is closed under subspace addition. Let $U_1, U_2 \in S_{W_0}$. If $U_2 \subseteq U_1$ or $U_1 \subseteq U_2$, the claim is trivially true. Thus, assume that $U_2 \nsubseteq U_1$, $U_1 \nsubseteq U_2$ and consider the following chain of inequalities.
\begin{eqnarray}
\mathcal{H}_N(U_1|W_0) & \stackrel{(a)}{\leq} & \mathcal{H}_N(U_1+U_2|U_1) \\
& \stackrel{(b)}{=} & \mathcal{H}_N(U_2|U_1) \\
& \stackrel{(c)}{\leq} & \mathcal{H}_N(U_2|U_1 \cap U_2) \nonumber \\
& \stackrel{(d)}{\leq} & \mathcal{H}_N(U_2|W_0), \label{eq:prop4_proof2}
\end{eqnarray}
where $(a)$ follows by applying Lemma \ref{lem:NE_2} to $W_0 \subsetneq U_1 \subsetneq U_1+U_2$ and using the fact that $\mathcal{H}_N(U_1|W_0) \leq \mathcal{H}_N(U_1+U_2|W_0)$ (since $U_1 \in S_{W_0}$), $(b)$ follows from Lemma \ref{lem:NE_1}, $(c)$ follows from Lemma \ref{lem:NE_3} and finally,
$(d)$ follows trivially, if $W_0 = U_1 \cap U_2$; else if $W_0 \subsetneq U_1 \cap U_2$, by applying Lemma \ref{lem:NE_2} to $W_0 \subsetneq U_1 \cap U_2 \subsetneq U_2$ and noting that $U_2 \in S_{W_0}$.

But, $\mathcal{H}_N(U_1|W_0) = \mathcal{H}_N(U_2|W_0)$ and thus all inequalities in \eqref{eq:prop4_proof2} are equalities. Especially, from $(a)$, we get that $\mathcal{H}_N(U_1+U_2|U_1)  = \mathcal{H}_N(U_1|W_0)$. Lemma \ref{lem:NE_2} now implies that $\mathcal{H}_N(U_1+U_2|W_0)  = \mathcal{H}_N(U_1|W_0)$ and thus $U_1+U_2 \in S_{W_0}$.

\section{Proof of Lemma \ref{lem:ss_ind_sources}} \label{app:ind_dist_subspace_chain}
Since $\{X_i\}$ and $\{Y_i\}$ are related via an invertible matrix, $< X_1, \ldots, X_m > = < Y_1, \ldots, Y_m >$. Thus, we will just find the subspace chain for the $\{Y_i\}$. Set $U_0 = \{ \underline{0} \}$ and $U_j = <Y_1, Y_2, \ldots, Y_{\sum_{i=1}^{j}\ell_i}>, \ 1 \leq j \leq r$. Let $U \subseteq V$ be such that $U_{j-1} \subsetneq U$. We will now show that $\mathcal{H}_N(U|U_{j-1}) \geq \mathcal{H}_N(U_j|U_{j-1})$ with equality only if $U \subseteq U_j$. This will imply that the chain $U_0 \subsetneq U_1 \subsetneq \ldots \subsetneq U_r$ satisfies the conditions of Theorem \ref{thm:chain_NCE} and hence, is the required chain.

Let $U\cap U_j = U_{j-1} \oplus A$ and $U = (U \cap U_j) \oplus B$, for some subspaces $A, B$. Then, $\mathcal{H}_N(U|U_{j-1})$ can be expanded as
\begin{eqnarray}
\mathcal{H}_N(U|U_{j-1}) & \stackrel{(a)}{=} & \frac{\mathcal{H}(A\oplus B|U_{j-1})}{\rho_A+\rho_B} \nonumber \\
& \stackrel{(b)}{=}    & \frac{\mathcal{H}(A|U_{j-1}) + \mathcal{H}(B|A \oplus U_{j-1})}{\rho_A+\rho_B} \nonumber \\
& \stackrel{(c)}{\geq} & \frac{\mathcal{H}(A|U_{j-1}) + \mathcal{H}(B|U_j)}{\rho_A+\rho_B} \nonumber \\
& \stackrel{(d)}{\geq} & \frac{ \rho_A H(Y_{\sum_{i=1}^{j-1}\ell_i+1}) + \rho_B  H(Y_{\sum_{i=1}^{j}\ell_i+1})}{\rho_A+\rho_B} \nonumber \\
& \stackrel{(e)}{\geq} & H(Y_{\sum_{i=1}^{j-1}\ell_i+1}) \ = \  \mathcal{H}_N(U_j|U_{j-1}),
\end{eqnarray}
where (a) follows from Lemma \ref{lem:NE_1}, (b), (c) both follow directly from the definition of conditional entropy in \eqref{eq:cond_entropy_subspace}, (d) follows from the fact that if $U'$ is a subspace such that $U' \cap U_j = \{ \underline{0} \}$, for any $j$, then
\begin{equation} \label{eq:lem_ind_sources1}
\mathcal{H}_N(U'|U_j) \geq H(Y_{\sum_{i=1}^{j}\ell_i+1}).
\end{equation}
This will be proved shortly. Finally, (e) follows from the assumption on the ordering of the entropies of $\{Y_i\}$ (see \eqref{eq:entropy_order}). Note that equality holds in (e) only if $B = \{ \underline{0} \}$.

We will now prove \eqref{eq:lem_ind_sources1}. Let $U' = < [Y_1, \ldots, Y_m]\Gamma_{U'} >$, for some $(m \times \rho_{U'})$ full rank matrix $\Gamma_{U'}$. Column reduce $\Gamma_{U'}$ by selecting, for any column, the last row which has a non zero entry and using that entry to make all the other entries in that row as zeros. Let $S = \{t_1, \ldots, t_{\rho_{U'}} \}$ denote the the row indices corresponding to the identity sub matrix which occur after the column reduction. Since $U' \cap U_j = \{ \underline{0} \}$, it must be true that
\begin{equation}
t_{i'} \ \geq \ {\sum_{i=1}^{j}\ell_i+1}, \ 1 \leq i' \leq \rho_{U'}. \label{eq:lem_indep_ss_prof_temp}
\end{equation}
Now, if we let $S^c = \{1,\ldots,m\} \backslash S$, we have
\begin{eqnarray}
\mathcal{H}_N(U'|U_j) & = & \frac{\mathcal{H}(U'|U_j)}{\rho_{U'}} \nonumber \\
& \stackrel{(a)}{\geq} & \frac{\mathcal{H}(U'|<Y_i,i \in S^c>)}{\rho_{U'}} \nonumber \\
& \stackrel{(b)}{=} & \frac{H(Y_{t_1},\ldots,Y_{t_{\rho_{U'}}})}{\rho_{U'}} \nonumber \\
& \stackrel{(c)}{\geq} & H(Y_{\sum_{i=1}^{j}\ell_i+1}),
\end{eqnarray}
where (a) follows since $U_j \subseteq <Y_i,i \in S^c>$, (b) follows from \eqref{eq:lem_indep_ss_prof_temp} and (c) follows from the assumption on the ordering of the entropies of $\{Y_i\}$ (see \eqref{eq:entropy_order}).

\section{Proof of Theorem \ref{thm:chain_SSA}} \label{app:proof_thm_chain_SSA}

The proof involves two steps, which are outlined next. Each step will be proved subsequently.

\vspace{0.05in}

\noindent \underline{Step $1$} : Consider the chain of subspaces $W^{(0)} \subsetneq W^{(1)} \subsetneq  \ldots \subsetneq W^{(r)}$ as obtained from Theorem \ref{thm:chain_NCE}. We will show that that the infimum of the achievable rates for decoding the subspace $W^{(j)}$ under the CC approach (see Section \ref{sec:CC}) is given by
\begin{align}
R_{\text{CC}}(W^{(j)}) & = \ \mathcal{H}_N \left( W^{(j)} | W^{(j-1)}\right), \ \forall \ 1 \leq j \leq r. \nonumber
\end{align}

\noindent  \underline{Step $2$} :  Next, consider any subspace $W \subseteq V$, such that $W \nsubseteq W^{(j-1)}$ and $W \subseteq W^{(j)}$. We show that an optimal subspace to decode $W$ under the SS approach is $W^{(j)}$, by showing that for any other subspace $W' \supseteq W$, we have
\begin{align}
R_{\text{CC}}(W') & \geq \ \mathcal{H}_N \left( W^{(j)} | W^{(j-1)}\right) \ = \ R_{\text{CC}}(W^{(j)}). \nonumber
\end{align}

\subsection{Proof of Step $1$}
Proof by induction on $j$. Statement follows for $j=1$, since from \eqref{eq:CC_inf}, we have
\begin{eqnarray}
R_{\text{CC}}(W^{(1)})  & = & \max_{U_1 \subsetneq W^{(1)}}  \mathcal{H}_N(W^{(1)}|U_1) \nonumber \\
& \stackrel{(a)}{=} & \mathcal{H}_N(W^{(1)}),
\end{eqnarray}
where $(a)$ follows by applying Lemma \ref{lem:NE_2} to $\{\underline{0}\} \subsetneq U_1 \subsetneq W^{(1)}$ and noting that from Theorem \ref{thm:chain_NCE} that $W^{(1)}$ has the least normalized entropy among all subspaces of $V$.
Now, assume that the statement is true for $j-1$, i.e.,
 \begin{align} \label{eq:induction_hyp}
\max_{U_1 \subsetneq W^{(j-1)}} \left\{ \mathcal{H}_N(W^{(j-1)}|U_1) \right \} & = \ \mathcal{H}_N(W^{(j-1)}|W^{(j-2)}).
\end{align}
Then, we need to prove that
\begin{align}
\max_{U_1 \subsetneq W^{(j)}} \left\{ \mathcal{H}_N(W^{(j)}|U_1) \right \} & = \ \mathcal{H}_N(W^{(j)}|W^{(j-1)}),
\end{align}
which will imply $R_{\text{CC}}(W^{(j)})  \  =  \ \mathcal{H}_N(W^{(j)}|W^{(j-1)})$. For any $U_1 \subsetneq W^{(j)}$, let $A \triangleq U_1 \cap W^{(j-1)}$ and let $A^c$ be its complement in $U_1$. Then
\begin{align}
\mathcal{H}(W^{(j)}|U_1) & \ \nonumber \\
& \hspace*{-.7in}= \ \mathcal{H}(W^{(j)}) - \mathcal{H}(A \oplus A^c) \nonumber \\
& \hspace*{-.7in}=\ \mathcal{H}(W^{(j-1)}) + \mathcal{H}(W^{(j)}|W^{(j-1)})- \mathcal{H}(A) - \mathcal{H}(A^c|A) \nonumber \\
& \hspace*{-.7in}\stackrel{(a)}{=} \ \mathcal{H}(W^{(j-1)}|A) + \mathcal{H}(W^{(j)}|W^{(j-1)}) - \mathcal{H}(A^c|A) \nonumber \\
&\hspace*{-.7in} \stackrel{(b)}{\leq} \ \mathcal{H}(W^{(j-1)}|A) + \mathcal{H}(W^{(j)}|W^{(j-1)}) - \mathcal{H}(A^c|W^{(j-1)}) \nonumber \\
&\hspace*{-.7in} \stackrel{(c)}{\leq} \ \mathcal{H}(W^{(j-1)}|A) + \mathcal{H}(W^{(j)}|W^{(j-1)}) - \rho_{A^c}\mathcal{H}_N(W^{(j)}|W^{(j-1)}) \nonumber \\
& \hspace*{-.7in}\stackrel{(d)}{\leq} \ (\rho_{W^{(j-1)}} - \rho_A)\mathcal{H}_N(W^{(j-1)}|W^{(j-2)}) \nonumber \\
&\hspace*{-.2in}+ \mathcal{H}(W^{(j)}|W^{(j-1)}) - \rho_{A^c}\mathcal{H}_N(W^{(j)}|W^{(j-1)}) \nonumber \\
&\hspace*{-.7in} \stackrel{(e)}{\leq} \ (\rho_{W^{(j-1)}} - \rho_A)\mathcal{H}_N(W^{(j)}|W^{(j-1)}) \nonumber \\
&\hspace*{-.2in}+ \mathcal{H}(W^{(j)}|W^{(j-1)}) - \rho_{A^c}\mathcal{H}_N(W^{(j)}|W^{(j-1)}) \nonumber \\
&\hspace*{-.7in} = \ \left(\rho_{W^{(j)}} - \rho_{U_1} \right) \mathcal{H}_N(W^{(j)}|W^{(j-1)}),
\end{align}
which implies that $\mathcal{H}_N(W^{(j)}|U_1) \leq \mathcal{H}_N(W^{(j)}|W^{(j-1)})$. Here,
$(a)$ and $(b)$ follow since $A \subseteq W^{(j-1)}$, $(c)$ follows trivially if $A^c = \{\underline{0}\}$; else from Theorem \ref{thm:chain_NCE}, $\mathcal{H}_N(W^{(j)}|W^{(j-1)}) < \mathcal{H}_N(A^c+W^{(j-1)}|W^{(j-1)}) = \mathcal{H}_N(A^c|W^{(j-1)})$. $(d)$ follows trivially if $A = W^{(j-1)}$; else by induction hypothesis on $j-1$ (put  $U_1 = A$ in \eqref{eq:induction_hyp}) and finally,
$(e)$ follows since  by Theorem \ref{thm:chain_NCE}, $\mathcal{H}_N(W^{(j-1)}|W^{(j-2)}) < \mathcal{H}_N(W^{(j)}|W^{(j-1)})$.

\subsection{Proof of Step $2$}
\vspace{-0.2in}
\begin{eqnarray}
R_{\text{CC}}(W') & = & \max_{U_1 \subsetneq W'} \left\{ \mathcal{H}_N(W'|U_1) \right\} \nonumber \\
& \stackrel{(a)}{\geq} &  \mathcal{H}_N(W'|W' \cap W^{(j-1)}) \nonumber \\
& \stackrel{(b)}{\geq} & \mathcal{H}_N(W'|W^{(j-1)}) \nonumber \\
& \stackrel{(c)}{=} & \mathcal{H}_N(W'+W^{(j-1)}|W^{(j-1)}) \nonumber \\
& \stackrel{(d)}{\geq} & \mathcal{H}_N(W^{(j)}|W^{(j-1)}) \label{eq:proof_step4_strict_ineq} \\
& \stackrel{(e)}{=} & R_{\text{CC}}(W^{(j)}),
\end{eqnarray}
where $(a)$ follows by substituting $U_1 = W' \cap W^{(j-1)}$, $(b)$ follows from Lemma \ref{lem:NE_3}, $(c)$ follows from Lemma \ref{lem:NE_1},
$(d)$ follows since by Theorem \ref{thm:chain_NCE}, $W^{(j)}$ is least normalized entropy subspace conditioned on ${W^{(j-1)}}$.
and finally, $(e)$ follows from Step 1.

\section{Existence of Nested Codes} \label{app:nested_codes}
For any $k_1 \leq k_2 \leq \ldots \leq k_j$, let $\bold{B}_{\ell}, \ell = 1, \ldots, j$, denote a random matrix uniformly picked from the
set of all $(k_{\ell} - k_{\ell-1}) \times n$ matrices over $\mathbb{F}_q$ (note here $k_0 = 0$). The encoding matrix for the
$\ell^{th}$ stage, $\bold{A}^{(\ell)}$, is assumed to have the nested form $\bold{A}^{(\ell)} = [ \bold{B}_1^t, \ldots,
\bold{B}_{\ell}^t]^t, \ 1 \leq \ell \leq j$. For any $\ell \leq j$, let $W^{(\ell)} = <{Y}_1, ..., {Y}_{\rho_{W^{(\ell)}}}>$.
As discussed in Section \ref{subsec:nested_codes}, the $\ell^{\text{th}}$ stage computes the complement of
${\bold{W}}^{(\ell-1)}$ in ${\bold{W}}^{(\ell)}$ using $\widehat{\bold{W}}^{(\ell-1)}$ (all the output up till the $(\ell -
1)^{\text{th}}$ stage) as side information. Now, for any fixed set of encoding matrices, let $\mathcal{E}_{\ell}$ denote the
error event up till the $\ell^{th}$ stage, i.e.,
\begin{equation}
\mathcal{E}_{\ell} \ : \ (\hat{\bold{Y}}_1, ..., \hat{\bold{Y}}_{\rho_{W^{(\ell)}}}) \neq ({\bold{Y}}_1, ..., {\bold{Y}}_{\rho_{W^{(\ell)}}}).
\end{equation}
Also let $P_{e, \ell}^{(n)}$ denote the probability of error in the $\ell^{th}$ stage assuming that all the previous stages were decoded correctly (i.e., when the $\ell^{th}$ stage receives $\bold{W}^{(\ell-1)}$ as side information). Thus, $P_{e, \ell}^{(n)} = P(\mathcal{E}_{\ell}|\mathcal{E}_{\ell-1}^c)$.  Thus the overall source averaged probability of error, $P_e^{(n)}$, in computing $\bold{W}^{(j)}$ can be upper bounded as
\begin{eqnarray}
P_e^{(n)} & = & P(\mathcal{E}_{j}) \label{eq:app_nested_codes_recur1}\\
& \leq & P(\mathcal{E}_{j-1}) + P(\mathcal{E}_{j}|\mathcal{E}_{j-1}^c) \\
& = & P(\mathcal{E}_{j-1}) + P_{e, j}^{(n)} \label{eq:app_nested_codes_recur2}\\
& \leq & \sum_{\ell = 1}^{j}P_{e, \ell}^{(n)},
\end{eqnarray}
where the last equation follows by repeating steps from \eqref{eq:app_nested_codes_recur1}-\eqref{eq:app_nested_codes_recur2}.
Averaging $P_e^{(n)}$ further over the ensemble of encoding matrices, we get
\begin{align}
<P_e^{(n)}> & =  \sum_{A^{(j)}}P_{\bold{A}^{(j)}}(A^{(j)})\sum_{\ell = 1}^{j}P_{e, \ell}^{(n)} \nonumber \\
& =  \sum_{\ell = 1}^{j}\sum_{A^{(\ell)}}P_{\bold{A}^{(\ell)}}(A^{(\ell)})P_{e, \ell}^{(n)},
\end{align}
where, in the last equation we have interchanged the order of the two summations and also used the fact that stage $\ell$
depends only on $\mathbf{B}_1, \ldots \mathbf{B}_{\ell}$ and hence $P_{\bold{A}^{(j)}}(A^{(j)})$ can be marginalized over the
incremental matrices of the remaining stages to get $P_{\bold{A}^{(\ell)}}(A^{(\ell)})$. But, now from the achievability
proof of Theorem \ref{thm:CC_side_info}, we know that the inside term
$\sum_{A^{(\ell)}}P_{\bold{A}^{(\ell)}}(A^{(\ell)})P_{e, \ell}^{(n)} \stackrel{n \rightarrow \infty}{\longrightarrow} 0$ if
\begin{equation}
 \frac{k_{\ell}}{n}\log{q}  >  \mathcal{H}_N \left(W^{(\ell)}|W^{(\ell-1)} \right), \forall  \ 1 \leq \ell \leq j.
\end{equation}
This proves the existence of the nested codes as claimed.

\end{document}